\pdfoutput=1
\documentclass[a4paper,11pt]{article}
\usepackage{jheppub} % For JHEP

\usepackage[T1]{fontenc}                           % for French font
\usepackage[utf8]{inputenc}                        % for non-ascii characters
\usepackage{microtype}                             % for spacing tweaks
\usepackage{amsmath,amssymb,mathtools}             % for maths
\usepackage{array,booktabs,multirow}               % for tables
\usepackage{lmodern}                               % for a vector font
\usepackage{etoolbox}                              % for programming
\usepackage[only, llbracket, rrbracket]{stmaryrd}  % for \llbracket and \rrbracket
\usepackage{relsize}                               % for smaller-sized acronyms

\usepackage{hyperref,bookmark}
\usepackage{tikz}
\usetikzlibrary{automata}
\usetikzlibrary{arrows}
\usetikzlibrary{calc}
\usetikzlibrary{decorations.markings}
\usetikzlibrary{decorations.pathreplacing}
\usetikzlibrary{intersections}
\usetikzlibrary{positioning}
\usetikzlibrary{topaths}
\usetikzlibrary{shapes.geometric}
\usetikzlibrary{shapes.misc}
\tikzset{->-/.style = {
    decoration = {markings, mark = at position #1 with {\arrow{stealth}}},
    postaction = {decorate}}}
\tikzset{color-group/.style = {
    shape = rounded rectangle,
    minimum size = 2.5ex,
    inner sep = .5ex,
    draw}}
\tikzset{flavor-group/.style = {
    shape = rectangle,
    minimum size = 2.5ex,
    inner sep = .5ex,
    draw}}
\tikzset{cf-group/.style = {
    shape = rounded rectangle,
    rounded rectangle right arc = none,
    draw}}
\tikzset{fc-group/.style = {
    shape = rounded rectangle,
    rounded rectangle left arc = none,
    draw}}
\tikzset{cross/.style={minimum width=1pt, path picture={
      \draw[black, very thick]
               (path picture bounding box.south east)
            -- (path picture bounding box.north west)
               (path picture bounding box.south west)
            -- (path picture bounding box.north east);
          }}}
\newcommand{\mathtikz}[2][]
  {\ensuremath{\vcenter{\hbox{%
          \begin{tikzpicture}[#1]#2\end{tikzpicture}}}}}
\newcommand{\quiver}[2][]
  {\mathtikz[semithick,node distance=3em,#1]{#2}}

\numberwithin{equation}{section}

\DeclareMathOperator*{\res}{res}

\let\Re\relax\let\Im\relax
\DeclareMathOperator{\Re}{Re}
\DeclareMathOperator{\Im}{Im}
\DeclarePairedDelimiter{\abs}{\lvert}{\rvert}
\DeclarePairedDelimiter{\vev}{\langle}{\rangle}
\DeclareMathOperator{\diag}{diag}
\DeclareMathOperator{\sign}{sign}
\DeclareMathOperator{\rank}{rank}

\newcommand{\I}    {{\mathrm{i}}}
\newcommand{\bbZ}  {{\mathbb{Z}}}

\newcommand{\bbC}  {{\mathbb{C}}}
\newcommand{\dd}[2][]{\mathop{\mathrm{d}#1#2}\nolimits}
\newcommand{\Nsusy}{{\mathcal{N}}}
\newcommand{\NToda}{N}
\newcommand{\dimToda}{\Delta}
\newcommand{\repr}{{\mathcal{R}}}

\newcommand{\intset}[2]{\llbracket #1, #2\rrbracket}
\newcommand{\anti}[1]{\texorpdfstring{{\protect\widetilde{#1}}}{#1'}}

\newcommand{\nomatrix}{\relax} % used as a marker
\def\gobblenomatrix#1\nomatrix{}
\newcommand{\bracketsorbmatrix}[1]{%
  \csname @\ifcat$\detokenize\expandafter{\gobblenomatrix#1\nomatrix}$first\else second\fi oftwo\endcsname
    {\!\begin{bmatrix}#1\end{bmatrix}\!}{[#1]}}
\newcommand{\Hypergeometric}[4][F]{%
  #1\left(\begin{smallmatrix}#2\\#3\end{smallmatrix}\middle|#4\right)}
\newcommand{\Fblockname}{{\mathcal{F}}}
\newcommand{\Fblock}[4]{%
  \mathinner{\Fblockname
    \ifstrempty{#1}{}{^{#1}}%
    \ifstrempty{#2}{}{_{#2}}%
    \ifstrempty{#3}{}{\bracketsorbmatrix{#3}}%
    \ifstrempty{#4}{}{\!(#4)}}}
\newcommand{\Braidingname}{{\mathbf{B}}}
\newcommand{\Braiding}[3]{%
  \Braidingname
  \ifstrempty{#1}{}{^{#1}}%
  \ifstrempty{#2}{}{_{#2}}%
  \ifstrempty{#3}{}{\bracketsorbmatrix{#3}}}
\newcommand{\Fusionname}{{\mathbf{F}}}
\newcommand{\Fusion}[3]{%
  \Fusionname
  \ifstrempty{#1}{}{^{#1}}%
  \ifstrempty{#2}{}{_{#2}}%
  \ifstrempty{#3}{}{\bracketsorbmatrix{#3}}}

\newcommand{\sch}{{\text{(s)}}}
\newcommand{\tch}{{\text{(t)}}}
\newcommand{\uch}{{\text{(u)}}}

\makeatletter
\let\oldfootnote\footnote
\renewcommand{\footnote}[1]{\oldfootnote{#1}\begingroup\futurelet\@let@token\my@footnote@test}
\newcommand{\my@footnote@test}{%
  \ifx\@let@token.\hskip-.2em\relax\fi
  \ifx\@let@token,\hskip-.2em\relax\fi\endgroup}
\makeatother

\newcommand{\USp}{\mathit{USp}}

\date{September 2019}
\title{S-duality wall of SQCD from Toda braiding}
\author{B. Le Floch}
\affiliation{Princeton Center for Theoretical Science, Princeton University, Princeton, NJ 08544, USA}
\affiliation{Philippe Meyer Institute, Physics Department, \'Ecole Normale Sup\'erieure, PSL Research University, 24 rue Lhomond, F-75231 Paris Cedex 05, France}
\emailAdd{bruno.le.floch@ens.fr}

\hypersetup{
  unicode,
  bookmarksnumbered,
  linktoc = all,
  pdfborderstyle = {/S/U/W 0.5},
  pdftitle = {S-duality wall of SQCD from Toda braiding},
  pdfauthor = {Bruno Le Floch}}

\abstract{%
  Exact field theory dualities can be implemented by duality domain walls such that
  passing any operator through the interface maps it to the dual operator.  This paper
  describes the S-duality wall of four-dimensional $\Nsusy=2$ $SU(N)$ SQCD with $2N$
  hypermultiplets in terms of fields on the defect, namely three-dimensional $\Nsusy=2$
  SQCD with gauge group $U(N-1)$ and $2N$ flavours, with a monopole superpotential.  The
  theory is self-dual under a duality found by Benini, Benvenuti and Pasquetti, in the
  same way that $T[SU(N)]$ (the S-duality wall of $\Nsusy=4$ super Yang--Mills) is
  self-mirror.  The domain-wall theory can also be realized as a limit of a $\USp(2N-2)$
  gauge theory; it reduces to known results for $N=2$.  The theory is found through the
  AGT correspondence by determining the braiding kernel of two semi-degenerate vertex
  operators in Toda CFT.}
\begin{document} 
\setcounter{tocdepth}{2}
\maketitle
\raggedbottom
%\flushbottom

\divide\hbadness by 3\relax \divide\vbadness by 3\relax % make sure we know about bad boxes

\bibliographystyle{alpha}

\section{Introduction}
\label{sec:intro}

Extended operators, such as Wilson and 't~Hooft loops \cite{Wilson:1974sk,tHooft:1977nqb},
surface operators \cite{Gukov:2006jk,Gukov:2008sn,Gukov:2014gja}, and domain walls
\cite{Gaiotto:2008sa,Gaiotto:2008sd,Gaiotto:2008ak} can serve as order parameters
\cite{Wilson:1974sk,tHooft:1977nqb,Gukov:2013zka} and help probe dualities of gauge
theories.  Exact all-scale dualities such as S-duality map all correlators of one theory
to the dual theory through some dictionary that must be worked out on a case by case
basis.  For instance, S-duality of four-dimensional $\Nsusy=2$ supersymmetric gauge
theories \cite{Seiberg:1994rs,Gaiotto:2009we} interchanges Wilson (electric) and 't~Hooft
(magnetic) loop operators
\cite{Kapustin:2005py,Alday:2009fs,Drukker:2009id,Drukker:2010jp}, as befitting of this
generalization of electric-magnetic duality.

Given an exact duality between theories $\mathcal{T}_1$ and~$\mathcal{T}_2$, what happens
if one dualizes one half-space, say $y\geq 0$?  Start with $\mathcal{T}_1$ on all of
space.  After dualizing, the system should be described by $\mathcal{T}_1$ for $y<0$ and
$\mathcal{T}_2$ for $y>0$, separated by a codimension~$1$ interface (wall) at $y=0$.  Only
a specific choice of interface can reproduce the physics of the original system.  For
instance, the position of the wall should not be observable so it must be topological away
from operator insertions.  When the interface passes through an operator inserted in the
region described by theory~$\mathcal{T}_2$, the operator should be replaced by its dual
description in theory~$\mathcal{T}_1$ to keep correlators fixed.  Such duality walls
\cite{Gaiotto:2008ak} are thus a compact way to encode how dualities act on operators
\cite{Fuchs:2007fw,Fuchs:2007tx,Sarkissian:2008dq,Kapustin:2009av,Tan:2009up,Tan:2013yza,Martucci:2014ema}.

The duality wall can often be described by coupling bulk fields of theories
$\mathcal{T}_1$ and $\mathcal{T}_2$ near the wall to additional degrees of freedom that
live only on the wall.  To determine what fields to put on the wall, one should decouple
the bulk theories by making their couplings weak.  This is impossible to do directly in
the most common case of a weak-strong duality, as making the bulk coupling small on one
side of the wall makes the other bulk theory strongly coupled.  To avoid this we begin
with a system described by theory $\mathcal{T}_1$ throughout space, with a Janus domain
wall, namely a $y$-dependent coupling that varies sharply near $y=0$ from some value to
another.  Now dualize the $y>0$ side to get theories $\mathcal{T}_1$ and $\mathcal{T}_2$
for $y\gtrless 0$, separated by a duality wall as before.  The couplings on the two sides
can now be sent to zero independently to decouple bulk dynamics from the theory on the
interface.  If enough is known about Janus walls the codimension~$1$ theory can be deduced
\cite{Gaiotto:2008ak,Hosomichi:2010vh,Terashima:2011qi,Terashima:2011xe,Dimofte:2011ju,Gadde:2013wq,Dimofte:2013lba,Gaiotto:2015usa,Gaiotto:2015una}.

For 4d $\Nsusy=4$ super Yang--Mills (SYM\@), $1/2$ BPS domain walls and boundary
conditions were extensively studied in \cite{Gaiotto:2008sa,Gaiotto:2008sd,Gaiotto:2008ak}
(and $1/4$ BPS ones in \cite{Hashimoto:2014vpa,Hashimoto:2014nwa}).  S-duality of
$\Nsusy=4$ SYM
changes the gauge group~$G$ to its Langlands dual ${}^LG$, and the S-duality wall is
described by gauging global symmetries $G\times {}^LG$ of a 3d $\Nsusy=4$ theory $T[G]$
using the 3d restrictions of the 4d vector multiplets on both sides of the wall.  The
analysis in~\cite{Gaiotto:2008ak} largely relies on
realizing $\Nsusy=4$ SYM as the world-sheet theory of $N$~D3~branes in the $G=SU(N)$ case
(other classical gauge groups are obtained using O3 planes), realizing boundary conditions
by ending D3~branes on D5~or NS5~branes and noting that S-duality of $\Nsusy=4$ SYM
descends from S-duality in IIB string theory.  The $T[SU(N)]$ theory is found to be given
by the following brane configuration from which one reads a quiver gauge theory
description:
\begin{equation}\label{intro-TSUN}
  \mathtikz[x=4em,y=5ex]{
    \foreach \B in {.0,.2,.4,.6}
      {
        \draw[dashed] (\B,-1) -- (\B,1.6);
        \draw (2\B,-1) -- (2\B,1.6);
        \draw (\B,\B) -- (2\B,\B);
      }
    \node at (0.3,-1.3) {\clap{D5(012789)}};
    \node at (1.3,-.3)  {\clap{D3(0126)}};
    \node at (2.3,-1.3) {\clap{NS5(012345)}};
  }
  \implies
  \quiver{
    \node (N)   [flavor-group]               {$N$};
    \node (N-1) [color-group, right of=N]    {$N-1$};
    \node (dots)[right=1em of N-1]  {$\cdots$};
    \node (2)   [color-group, right of=dots] {2};
    \node (1)   [color-group, right of=2]    {1};
    \draw (N) -- (N-1);
    \draw (N-1) -- (dots);
    \draw (dots) -- (2);
    \draw (2) -- (1);
  }
\end{equation}
More precisely, $T[SU(N)]$ is the infrared limit of the gauge theory with gauge group
$U(N-1)\times U(N-2)\times\dots\times U(1)$ and hypermultiplets in the bifundamental
representation of each pair $U(k)\times U(k-1)$, as well as $N$~hypermultiplets in the
fundamental representation of $U(N-1)$, which have an $SU(N)$ flavour symmetry.  Since IIB
S-duality interchanges D5 and NS5 branes, leaving the brane diagram essentially invariant,
$T[SU(N)]$ is self-mirror so its $U(1)^{N-1}$ topological symmetry enhances to
another\footnote{Throughout we will ignore global aspects of flavour symmetries: here one
  $SU(N)$ should be $PSU(N)$.} $SU(N)={}^LSU(N)$.  However,
this construction through IIB strings does not apply to S-duality of 4d $\Nsusy=2$
theories\footnote{One exception is that the world-volume theory of fractional D3 branes at
  the center of an orbifold $\mathbb{R}^4/\mathbb{Z}_k$ is a 4d $\Nsusy=2$ necklace
  quiver.  However, IIB string theory only provides $SL(2,\mathbb{Z})$ S-dualities acting
  on all gauge couplings at once, while there exists a much richer set of dualities acting
  ``locally'' on the quiver.}.

For the large class of 4d $\Nsusy=2$ $SU(N)$ gauge theories obtained by twisted
dimensional reduction of 6d $(2,0)$ $A_{N-1}$ superconformal theory on a punctured Riemann
surface~$\Sigma$, the AGT correspondence provides another handle~\cite{Alday:2009aq} (see
also~\cite{Wyllard:2009hg,Hama:2012bg,Nosaka:2013cpa}).  It relates observables of the 4d
theory on an ellipsoid~$S^4_b$ to observables of the Toda CFT on~$\Sigma$, a
generalization of the Liouville CFT with a larger symmetry algebra~$W_N$.  In particular,
an S-duality domain wall placed along the equator (or a parallel) $S^3_b$ of~$S^4_b$
corresponds to a certain $W_N$~braiding transformation~\cite{Drukker:2010jp}.  From the
integral kernel that braids certain $W_N$~primary operators of the Toda CFT one can thus
deduce the partition function of the S-duality domain wall, hence the $S^3_b$ partition
function of its description as a 3d gauge theory, as we explain in \autoref{sec:AGT}.  One
is then left with finding a 3d gauge theory with the given $S^3_b$ partition function.
This strategy was applied in~\cite{Hosomichi:2010vh} to the mass deformation of 4d
$\Nsusy=4$ and in~\cite{Teschner:2012em} to 4d $\Nsusy=2$ $SU(2)$ SQCD with
$4$~fundamental hypermultiplets.

While we concentrate on 4d $\Nsusy=2$ $SU(N)$ SQCD with $2N$ fundamental hypermultiplets,
the 3d $\Nsusy=2$ description of the SQCD S-duality wall that we give momentarily also
implements S-duality of any balanced $SU(N)$ gauge node in a 4d $\Nsusy=2$ quiver (see the
conclusion section).  We have not worked out the case of $N_f<2N$ hypermultiplets, but
it should simply amount to making some hypermultiplets massive.

We determine the relevant braiding kernel~\eqref{kernel-expression} in
\autoref{sec:kernel} as a continuation of braiding matrices~\eqref{matrices-sym-expr}
worked out in \autoref{sec:matrices} using 2d CFT pentagon relations.  Comparing it to
explicit expressions of~$S^3_b$ partition functions from localization
\cite{Jafferis:2010un,Hama:2010av,Hama:2011ea} we describe in \autoref{sec:wall} the
S-duality wall of 4d $\Nsusy=2$ $SU(N)$ SQCD as 3d $\Nsusy=2$ $U(N-1)$ SQCD with a
monopole superpotential on the wall, coupled to the 4d theories on both sides of the wall.
We then find evidence that the 3d theory can also be obtained as a certain limit of an
$\USp(2N-2)$ theory, in which some symmetries are more manifest.  Altogether,
\begin{equation}\label{intro-wallZZ}
  \vev{\text{S-duality wall}}
  =
  Z \!\left[
    \quiver{
      \node (1)   [color-group] {$\scriptstyle U(N-1)$};
      \node (R)   [flavor-group, right=1em of 1] {$2N$};
      \node (A)   [color-group, above=3ex of 1] {$N$};
      \node (B)   [color-group, below=3ex of 1] {$N$};
      \draw[->-=.55] (R) -- (1);
      \draw[->-=.7] (1) -- (A);
      \draw[->-=.7] (1) -- (B);
      \draw (A) -- (R) node [midway, above right=-4pt] {4d};
      \draw (B) -- (R) node [midway, below right=-4pt] {4d};
    }
  \right]
  =
  \lim_{\mu\to\pm\infty}
  Z \!\left[
    \quiver{
      \node (1)   [color-group] {$\scriptstyle \USp(2N-2)$};
      \node (R)   [flavor-group, right=2.5em of 1] {$2N$};
      \node (A)   [color-group, above=3ex of 1] {$N$};
      \node (B)   [color-group, below=3ex of 1] {$N$};
      \draw[->-=.55] (R) -- (1) node [pos=.7,below=-.8ex] {$-\mu$};
      \draw[->-=.7] (1) -- (A) node [midway,left] {$+\mu$};
      \draw[->-=.7] (1) -- (B) node [midway,left] {$+\mu$};
      \draw (A) -- (R) node [midway, above right=-4pt] {4d};
      \draw (B) -- (R) node [midway, below right=-4pt] {4d};
    }
  \right] \,.
\end{equation}
In these quiver, the upper and lower round nodes labelled~$N$ denote $SU(N)$ gauge groups
of the 4d theories on both sides of the wall, the diagonal edges stemming from them are
$2N$ hypermultiplets (again on both sides of the wall) that share a common $SU(2N)$
flavour symmetry across the wall.  The rest of the quiver describes the 3d $\Nsusy=2$
theory on the wall: a $U(N-1)$ or $\USp(2N-2)$ vector multiplet coupled to $2N$
fundamental and $N+N$ antifundamental chiral multiplets.  The labels $\pm\mu$ will be
explained shortly.  Additionally, the 3d~and 4d~matter multiplets are coupled by a cubic
superpotential~$W$ along the defect\footnote{Version~1 of this paper was lacking $V_++V_-$
  and $Y$ so the proposed 4d/3d system had too many symmetries.  Thanks to Sara Pasquetti
  for pointing this out.  In fact, the corrected 3d theory with $W=V_++V_-$ is self-dual:
  see \cite{Dimofte:2012pd} for $N=2$ and \cite{Benini:2017dud} (after our version~1)
  for all~$N$.  This duality turns out to make the $\USp(2N-2)$ theory unnecessary for our
  purposes, but we keep it as an interesting variant.}:
\begin{equation}\label{intro-supo}
  W = \sum_{f=1}^{2N} \sum_{s=1}^{N} \Bigl( \Phi_{fs}\bigr|_{\text{3d}} \anti{q}_{s} q_{f} + \Phi'_{fs}\bigr|_{\text{3d}} \anti{q}'_{s} q_{f} \Bigr) +
  \begin{cases}
    V_+ + V_- & \text{for the $U(N-1)$ theory,} \\
    Y & \text{for the $\USp(2N-2)$ theory.}
  \end{cases}
\end{equation}
Here, $\anti{q}_s$ and $\anti{q}'_{s}$ denote the $2N$ antifundamental chiral multiplets
and $q_{f}$ the $2N$ fundamental ones, while $\Phi|_{\text{3d}}$ and~$\Phi'|_{\text{3d}}$
are limits of 4d hypermultiplets at the interface.  Following standard notations, $V_{\pm}$
are monopole operators with $U(1)_T$ charges $\pm 1$ in the $U(N-1)$ theory, while for the
$\USp(2N-2)$ theory the monopole is denoted by~$Y$ (see, e.g., \cite{Aharony:1997gp}).
Coefficients of the superpotential cannot be determined by our methods.

The rest of this paper is organized as follows.  \autoref{sec:AGT} explains the AGT
correspondence relating the S-duality domain wall of 4d $\Nsusy=2$ $SU(N)$ SQCD to a $W_N$
braiding kernel\footnote{While the braiding kernel is only sensitive to the chiral
  symmetry algebra $W_N$ we will sometimes call it Toda CFT braiding kernel
  because Toda CFT is the prime example of $W_N$-symmetric theory.}  and gives an
introduction to the Toda CFT\@.  \autoref{sec:matrices} evaluates some braiding
matrices~\eqref{matrices-sym-expr} which are special cases of the braiding kernel when an
operator is fully degenerate.  \autoref{sec:kernel} generalizes these discrete results to
a continuous braiding kernel~\eqref{kernel-expression}.  It describes how this integral
kernel reduces to Liouville CFT braiding kernels~\cite{Ponsot:1999uf,Ponsot:2000mt} and proves that the
kernel obeys a shift relation deduced from Moore--Seiberg pentagon
identities~\cite{Moore:1988uz}.  A reader with no interest in Toda CFT can jump directly
to \autoref{sec:wall}, where we extract the above gauge theory description of the
S-duality wall from the braiding kernel and explain how continuous flavour symmetries and
various $\mathbb{Z}_2$ symmetries (including self-dualities) match between the different
descriptions.  We conclude in \autoref{sec:conclusion} with some remarks on class~S quiver
gauge theories.

\section{AGT relation}
\label{sec:AGT}

This section describes supersymmetric localization for 4d $\Nsusy=2$ theories, then
translates S-duality domain walls of $SU(N)$ SQCD to the 2d CFT language using the AGT
correspondence.

\subsection{Localization}
\label{ssec:AGT-loc}

For a given choice of supercharge~$\mathcal{Q}$ under which a path integral is invariant,
localization reduces the path integral to a simpler integral over $\mathcal{Q}$-invariant
field configurations only\footnote{More precisely, localization relies on a choice of
  $\mathcal{Q}$-closed and $\mathcal{Q}$-exact (and positive) deformation term $\delta S$
  for the action: the deformed path integral including a factor $\exp(-t\delta S)$ is then
  independent of $t\geq 0$.  At large~$t$ the saddle point approximation is exact.  For an
  appropriate~$\delta S$, saddle points are $\mathcal{Q}$-invariant configurations.}.  As
an IR cutoff we place the theory on the ellipsoid~$S^4_b$ given in coordinates as
\begin{equation}
  x_0^2 + b^{-2} (x_1^2+x_2^2) + b^2 (x_3^2+x_4^2) = 1 \,.
\end{equation}
Preserving supersymmetry requires non-trivial background fields~\cite{Hama:2012bg}.  There
exists another deformation of the sphere in which the parameter~$b$ is complex rather than
real~\cite{Nosaka:2013cpa}.  Domain walls will be placed at constant~$x_0$, such as the
equator~$S^3_b$ at $x_0=0$.

The partition function of 4d $\Nsusy=2$ theories on~$S^4_b$ was
computed~\cite{Pestun:2007rz,Hama:2012bg} using a supercharge~$\mathcal{Q}$ whose square
rotates the ellipsoid in the planes $(x_1,x_2)$ and $(x_3,x_4)$ and leave two poles
$x_0=\pm 1$ fixed.  In $\mathcal{Q}$-invariant configurations one vector multiplet scalar
takes a constant value~$\I a$ and other fields vanish, except at the North and South poles
where there can be instantons and anti-instantons.  The exact $S^4_b$~partition function
is then
\begin{equation}
  Z_{S^4_b} = \int \dd{a} e^{-S_{\text{cl}}(a,x,\bar{x})} Z_{\text{1-loop}}(m,a) Z_{\text{inst}}(m,a,x) Z_{\text{anti-inst}}(m,a,\bar{x})
\end{equation}
where the real scalar $a$~runs over the gauge Lie algebra, $x$ stands for exponentiated
gauge coupling constants and $m$ for the masses.  The integrand consists of a classical
contribution $\exp(-S_{\text{cl}})$ evaluated at the saddle point (away from the poles), a
one-loop determinant due to fluctuations around the saddle point, and (anti)\-instanton
contributions at the poles, which depend (anti)\-holomorphically on~$x$.  These instanton
partition functions are power series that normally converge in a finite region near the
weakly-coupled limit $x\to 0$.  After reducing the integral to the Cartan algebra,
combining the resulting Vandermonde determinant with $Z_{\text{1-loop}}$, and factorizing
the classical contribution into holomorphic functions of $x$ and~$\bar{x}$, one
gets~\eqref{AGT-ZS4} below.
S-dual theories have equal ellipsoid partition functions, hence
\begin{align}
  \label{AGT-ZS4}
  Z_{S^4_b} & = \int \dd{a} C(m,a) f(m,a,x) f(m,a,\bar{x})
  \\
  \label{AGT-ZS4dual}
          & = \int \dd{a_{D}} C_{D}(m,a_{D}) f_{D}(m,a_{D},x_{D}) f_{D}(m,a_{D},\bar{x}_{D}) \,.
\end{align}
While the global symmetries and masses~$m$ of the two 4d theories are shared up to permutations, their matter
content and coupling constants may be very different.  For a fixed $\bar{x}$ (and
$\bar{x}_{D}$) the two integral representations express~$Z$ in different
bases of holomorphic functions $f$ and~$f_{D}$, hence there should exist a change of basis
$B(m,a,a')$, called S-duality kernel:
\begin{equation}\label{AGT-Sduality}
  f(m,a,x) = \int \dd{a'} B(m,a,a') f_{D}(m,a',x_{D}) \,.
\end{equation}

Let us repeat the localization computation, but with a Janus domain wall that
interpolates between $x$ in the North half-ellipsoid and $x'$ in the South half-ellipsoid.
While the $\mathcal{Q}$-invariant field configurations are unaffected by the wall, the
(anti)instanton contributions from the North and South poles are evaluated using $x$ and
$\bar{x}'$, respectively.  The classical contribution is affected in a similar way
\cite[Section 5]{Drukker:2010jp}, and one gets
\begin{equation}
  \vev{\text{Janus wall}} = \int \dd{a} C(m,a) f(m,a,x) f(m,a,\bar{x}') \,.
\end{equation}
When the couplings are S-dual ($\bar{x}'_{D}=\bar{x}$), one can apply
S-duality~\eqref{AGT-Sduality} to the South half-ellipsoid and obtain a duality wall:
\begin{equation}\label{AGT-Swallvev}
  \vev{\text{S-duality wall}}
  = \int \dd{a} \int \dd{a'} C(m,a) B(m,a,a')
  f(m,a,x) f_{D}(m,a',\bar{x})
  \,.
\end{equation}

On the other hand, consider the 4d $\Nsusy=2$ theory on~$S^4_b$ with coupling~$x$
throughout the ellipsoid, coupled to 3d fields on the equator.  $\mathcal{Q}$-invariant
configurations of such a 4d/3d system have constant vector multiplet scalars $\I a$
and~$\I a'$ on the two halves of~$S^4_b$, instantons at each pole, and possibly additional
non-trivial field configurations on the 3d equator.  The contribution from 3d fields is
simply the $S^3$~partition function of the theory obtained by freezing the 4d fields to
their values $(\I a,\I a')$.  The 4d classical and instanton contributions combine as
before into $f(m,a,x)$ and $f_{D}(m,a',\bar{x})$ (provided an appropriate Chern-Simons
term is included on the wall), and finally 4d one-loop determinants on each half-ellipsoid
are functions of $(m,a)$ and of $(m,a')$.  Altogether, \eqref{AGT-Swallvev}~can be
reproduced provided one finds a 3d theory whose ellipsoid partition function is
essentially $B(m,a,a')$:
\begin{equation}\label{AGT-ZS3-CBZZ}
  Z_{S^3_b}(m,a,a') = \frac{C(m,a) B(m,a,a')}{Z_{\text{1-loop}}^{\text{half-ellipsoid}}(m,a) Z_{\text{1-loop}}^{\text{half-ellipsoid}}(m,a')} \,.
\end{equation}
In particular, the 3d partition function should depend on 4d masses and vector multiplet
scalars $(m, a, a')$ which lie in the Cartan algebras of the global symmetry group (shared
by the two 4d theories) and of both gauge groups.  The 3d theory must therefore contain
fields charged under each of these symmetries.  For instance, the domain wall theory
$T[SU(N)]$ of 4d $\Nsusy=4$ $SU(N)$ SYM~\eqref{intro-TSUN} has a (non-manifest) global
symmetry $U(1)\times SU(N)^2$.

We are interested in 4d $\Nsusy=2$ $SU(N)$ SQCD with $2N$ flavours.  Let $m_f$
($f=1,\ldots,2N$) be the masses of these hypermultiplets.  S-duality inverts the $U(1)$
subgroup of the $U(2N)$ flavour symmetry: namely masses in the dual theory are $m_f-2m$,
where $m=\frac{1}{2N}\sum_f m_f$.  The 3d gauge theory description, which we find in
\autoref{sec:wall}, must thus have the flavour symmetry
$U(1)\times SU(2N)\times SU(N)\times SU(N)$ (up to discrete factors).
This symmetry is explicitly broken by $(m,a,a')$.

To find this 3d gauge theory description we must evaluate the right-hand side
of~\eqref{AGT-ZS3-CBZZ}.  The relevant structure constants~$C(m,a)$ are known.  The one-loop
determinants of 4d vector and hypermultiplets are only known on the full ellipsoid and not
on a half-ellipsoid.  We take as inspiration the analoguous situation in two dimensions:
sphere one-loop determinants involve a combination $\Gamma(x)/\Gamma(1-x)$ while
hemisphere ones involve $\Gamma(x)$ or $1/\Gamma(1-x)$ depending on boundary
conditions~\cite{Sugishita:2013jca,Honda:2013uca,Hori:2013ika}.  In four dimensions, we
note that the one-loop determinant of a hypermultiplet of mass~$m$
is\footnote{See \autoref{ssec:functions} for properties of special functions.}
$\Gamma_b\bigl(\frac{b+b^{-1}}{2}+\I m\bigr) \Gamma_b\bigl(\frac{b+b^{-1}}{2}-\I
m\bigr)$.  It is thus natural to propose that
the one-loop determinant on a hemisphere (or rather a half-ellipsoid) is
a single one of these two $\Gamma_b$~functions\footnote{This was
  previously suggested for instance in \cite[page 24]{Dimofte:2012pd}
  and \cite[page 29]{Bullimore:2013xsa}.}.  The situation is similar for
vector multiplets.  Even if the proposal turns out to be incorrect, our main statement
identifying what 3d theory to couple to the two 4d theories will hold, since one-loop
determinants depend on $a$ and~$a'$ separately.  The most important ingredient
in~\eqref{AGT-ZS3-CBZZ} is the S-duality kernel~$B$, which mixes $a$ and~$a'$.

\subsection{Toda CFT}
\label{ssec:AGT-Toda}

To determine the kernel~$B$ we will use the AGT correspondence~\cite{Alday:2009aq}, found
by remarking that the various factorizations \eqref{AGT-ZS4} and~\eqref{AGT-ZS4dual} of
$Z$~into holomorphic factors are reminiscent of conformal block decompositions in 2d
CFT\@.  Observables of 4d $\Nsusy=2$ $SU(N)$ gauge theories (of class~S) on $S^4_b$ are
equal to observables in the $A_{N-1}$ Toda CFT\@.

This
generalization of the Liouville CFT (the case $N=2$) has a symmetry algebra~$W_N$ (for
$N=2$, the Virasoro algebra).  We use standard notations\footnote{Let $h_1,\ldots,h_N$
  denote the weights of the fundamental representation of~$A_{N-1}$, which sum to~$0$.
  These form an overcomplete basis of the Cartan algebra~$\mathfrak{h}$, identified
  to~$\mathfrak{h}^*$ by the Killing form defined by $\vev{h_i,h_j}=\delta_{ij}-1/N$.  The
  highest weight of any representation of~$A_{N-1}$ can be written
  $\Omega=n_1h_1+\cdots+n_N h_N$ with integers $n_1\geq\cdots\geq n_N=0$.  We also let
  $\rho=\frac{1}{2}\sum_{i<j}^{N} (h_i-h_j) = \sum_{i=1}^{N} \frac{N+1-2i}{2} h_i$ be the
  half-sum of positive roots and $Q=(b+b^{-1})\rho$.}.  The vertex
operators~$\widehat{V}_{\alpha}$, whose momentum~$\alpha$ depends on $N-1$ parameters, are
$W_N$~primaries of dimension $\dimToda(\alpha)=\frac{1}{2}\vev{\alpha,2Q-\alpha}$.  They
are invariant under Weyl transformations, which permute their components
$\vev{\alpha-Q,h_i}$\footnote{Momenta are elements of~$\mathfrak{h}$.  Normalizability
  requires $\alpha-Q$ to be an imaginary element of~$\mathfrak{h}$.  The dimension of
  $V_{\alpha}$ and other quantum numbers are invariant under Weyl transformations, and it
  turns out that the Toda CFT has a single operator with a given set of quantum numbers,
  hence $V_{\alpha}$~itself is invariant up to a scalar.  We choose a
  normalization~$\widehat{V}_{\alpha}$ invariant under Weyl transformations; the
  normalization also avoids annoying constants in the AGT relation, and it does not affect
  conformal blocks nor braiding.}.  Among these primary operators we call the
one-parameter class of momenta $\alpha=\kappa h_1$ semi-degenerate, or
simple\footnote{Many authors consider semi-degenerate momenta of the form
  $\alpha=-\lambda h_N$.  This choice is mapped to ours by a Weyl reflection:
  $-\lambda h_N\to (N(b+b^{-1})-\lambda)h_1$.}, and the discrete set $\alpha=-Kbh_1$ (for integer
$K\geq 0$) fully degenerate\footnote{More generally, $\alpha=-b\Omega_1-b^{-1}\Omega_2$,
  where $\Omega_1$ and~$\Omega_2$ are highest weights of two representations of~$A_{N-1}$,
  are fully degenerate and form a discrete set.  We will always take $\Omega_2=0$ and
  $\Omega_1=Kh_1$, highest weight of the $K$-th symmetric representation of~$A_{N-1}$.
  Both semi-degenerate and fully degenerate primary operators are interesting because
  there are null-vectors among their $W_N$~descendants.}.

The $S^4_b$~partition function of 4d $\Nsusy=2$ SQCD is then equal to a sphere correlator
of two generic and two simple vertex operators:
\begin{equation}\label{AGT-ZS4-VVVV}
  Z_{S^4_b}(\text{SQCD})
  =
  \abs{x}^{2\gamma_0} \abs{1-x}^{2\gamma_1}
  \vev{\widehat{V}_{\alpha_3}(\infty) \widehat{V}_{\kappa_4 h_1}(1) \widehat{V}_{\kappa_2 h_1}(x,\bar{x}) \widehat{V}_{\alpha_1}(0)}
\end{equation}
The cross-ratio~$x$ of their positions is the exponentiated gauge coupling, mapped to
$1/x$ by S-duality.  The unimportant exponents~$\gamma_i$ can be fixed by matching Toda
CFT and gauge theory asymptotics as $x\to 0,\infty$.  The momenta $\alpha_1$ and
$\kappa_2 h_1$ encode $N$~hypermultiplet masses and $\alpha_3$ and $\kappa_4 h_1$ the
other~$N$.  The Toda CFT correlator thus does not make all gauge theory symmetries
explicit, which leads to various sign asymmetries.  Momenta are
\begin{equation}\label{AGT-dict}
  \begin{aligned}
    \alpha_1 & = Q + \sum_{j=1}^{N} ( \I m_j h_j ) &
    \kappa_2 & = N \biggl( \frac{b+b^{-1}}{2} - \frac{1}{N} \sum_{j=1}^{N} \I m_j \biggr) \\
    \alpha_3 & = Q - \sum_{j=1}^{N} ( \I m_{j+N} h_j ) \qquad &
    \kappa_4 & = N \biggl( \frac{b+b^{-1}}{2} + \frac{1}{N} \sum_{j=1}^{N} \I m_{j+N} \biggr) \,.
  \end{aligned}
\end{equation}

From the Toda CFT point of view, the two S-dual decompositions \eqref{AGT-ZS4}
and~\eqref{AGT-ZS4dual} of the partition function are conformal block decompositions
obtained by taking the operator product expansion (OPE) of
$\widehat{V}_{\kappa_2 h_1}(x,\bar{x})$ with either $\widehat{V}_{\alpha_1}(0)$
(``s-channel decomposition'') or with $\widehat{V}_{\alpha_3}(\infty)$ (``u-channel
decomposition'').  Considering the s-channel for definiteness, the integration variable
$a$ parametrizes the primary operator resulting from the OPE of $\widehat{V}_{\alpha_1}$
with~$\widehat{V}_{\kappa_2 h_1}$.  The one-loop contributions $C(m,a)$ are structure
constants of the Toda CFT, and the (anti)instanton partition functions $f(m,a,x)$ and
$f(m,a,\bar{x})$ are (anti)holomorphic s-channel conformal blocks.

Four-point function of $W_N$~primary operators do not decompose so simply into
(anti)holomorphic conformal blocks in general.  Inserting a complete set of states (both
primaries and their descendants) in a generic four-point function $\vev{V_4V_3V_2V_1}$
gives schematically
$\int\dd{\alpha}\sum_{I,\overline{I}}
\vev{V_4V_3(W_{-I}\overline{W}_{-\overline{I}}V_{\alpha})}
\vev{(W_{-I}\overline{W}_{-\overline{I}}V_{2Q-\alpha})V_2V_1}$ where $W_{-I}$ and
$\overline{W}_{-\overline{I}}$ are an orthonormal basis of the left/right-moving
$W_N$~algebras and we used $\vev{V_{\alpha}V_{2Q-\alpha}}=1$.  When each three-point
function feature a semi-degenerate vertex operator, its null-vectors can be used to
convert the action of $W$ and~$\overline{W}$ into that of Virasoro generators only, which
are known to act as differential operators.  For the four-point
function~\eqref{AGT-ZS4-VVVV} of interest the s-channel and u-channel have this property
(not the t-channel), which guarantees that the contributions from descendants of a given
primary operator factorize into conformal blocks.

The s-channel and u-channel decompositions converge in different regions
$\abs{x}\lessgtr 1$.  In both cases, every conformal block is a fractional power of~$x$
multiplied by a power series which converges in the unit disk.  They can in fact be
analytically continued to the whole plane minus branch cuts joining $0$, $1$,
and~$\infty$.  A convenient choice will be to cut along $[0,\infty)$, in other words
normalize conformal blocks so that their leading term is a fractional power of $(-x)$.  We
define the braiding kernel as the integral kernel expressing s-channel blocks in terms of
u-channel ones after analytic continuation.  This is the Toda CFT translation of the
S-duality kernel defined in~\eqref{AGT-Sduality}.  For comparison we write formulas next
to each other:
\begin{align}
  f(m,a,x) & = \int \dd{a'} B(m,a,a') f_{D}(m,a',x_{D}) \\
  \Fblock{}{}
    {
      \mathtikz[x=1em,y=1ex,thick]{
        \draw[->-=.55] (0,0) -- (2,0);
        \draw[->-=.55] (6,0) -- (4,0);
        \draw[->-=.55] (4,0) -- (2,0);
        \draw[->-=.55] (2,-4) -- (2,0);
        \draw[->-=.55] (4,4) -- (4,0);
        \node at (0.3,1.5) {$\alpha_3$};
        \node at (5.7,-1.5) {$\alpha_1$};
        \node at (3,1.5) {$\alpha_{12}$};
        \node at (5.3,3) {$\kappa_2 h_1$};
        \node at (0.7,-3) {$\kappa_4 h_1$};
      }
    }{x}
  &
  = \int \dd{\alpha_{32}}
    \Braiding{}{\alpha_{12}\alpha_{32}}{\kappa_4 h_1&\kappa_2 h_1\\\alpha_3&\alpha_1}
  \Fblock{}{}
    {
      \mathtikz[x=1em,y=1ex,thick]{
        \draw[->-=.55] (0,0) -- (2,0);
        \draw[->-=.55] (6,0) -- (4,0);
        \draw[->-=.55] (2,0) -- (4,0);
        \draw[->-=.55] (2,4) -- (2,0);
        \draw[->-=.55] (4,-4) -- (4,0);
        \node at (0.3,-1.5) {$\alpha_3$};
        \node at (5.7,1.5) {$\alpha_1$};
        \node at (3,-1.5) {$\alpha_{32}$};
        \node at (5.3,-3) {$\kappa_4 h_1$};
        \node at (0.7,3) {$\kappa_2 h_1$};
      }
    }{x}
\end{align}
The $\alpha_{32}$ integral runs over imaginary values for $\alpha_{32}-Q$.  The blocks
$\Fblock{}{}{}{x}$ and $f(x)$ differ by the same factor $(-x)^{\gamma_0}(1-x)^{\gamma_1}$
in both channels, hence $B(m,a,a')=\Braiding{}{\alpha_{12}\alpha_{32}}{}$.

In the normalization where the leading term of $f(x)$ is a power of~$x$, the S-duality
kernel is changed by a phase due to altered branch cuts.  Using
$x^{\lambda}=e^{\epsilon\I\pi\lambda}(-x)^{\lambda}$ with $\epsilon=\sign\Im x$, and given
the semi-classical limits $f(m,a,x)\sim x^{\frac{1}{2}\sum_j a_j^2}$ and
$f(m,a,x)\sim x^{-\frac{1}{2}\sum_j a_j'^2}$ as $x\to 0,\infty$, the S-duality kernel in
this normalization is
\begin{equation}\label{AGT-phase}
  B^\epsilon(m,a,a') = e^{\epsilon\I\pi[\frac{1}{2}\sum_j a_j^2+\frac{1}{2}\sum_j a_j'^2]} B(m,a,a') \,.
\end{equation}
We will interpret these phases as a Chern--Simons term on the wall.  The braiding kernel
receives similar phases.

Our goal is to find the kernel $\Braiding{}{\alpha_{12}\alpha_{32}}{}$.  We determine it
in \autoref{sec:matrices} when $\kappa_2 h_1$ is fully degenerate, namely $\kappa_2=-Kb$
with $K\geq 0$ an integer, then generalize to all~$\kappa_2$ in \autoref{sec:kernel}.

\subsection{\label{ssec:functions}Special functions}

Besides the Barnes double-Gamma function $\Gamma_b=\Gamma_{1/b}$, normalized by
$\Gamma_b(\frac{b+b^{-1}}{2})=1$, we need the double-Sine function
$S_b(x) = S_{1/b}(x) = \Gamma_b(x) / \Gamma_b(b+b^{-1}-x) = 1/S_b(b+b^{-1}-x)$, and the
Upsilon function
$\Upsilon_b(x) = \Upsilon_{1/b}(x) = 1 / (\Gamma_b(x)\Gamma_b(b+b^{-1}-x)) =
\Upsilon_b(b+b^{-1}-x)$.  All are meromorphic in~$x$ and have the following zeros, poles
and shift relations expressed using $\gamma(y)=\Gamma(y)/\Gamma(1-y)$ and the Euler
identity $\Gamma(y)\Gamma(1-y)=\pi/\sin(\pi y)$.  In addition, $S_b(b)=b$.
\begin{center}
  \begin{tabular}{cccr@{}l}
    \toprule
    & $-b\mathbb{Z}_{\geq 0}-\frac{1}{b}\mathbb{Z}_{\geq 0}$ & $b\mathbb{Z}_{\geq 1}+\frac{1}{b}\mathbb{Z}_{\geq 1}$ & \multicolumn{2}{c}{Shift relation} \\
    \midrule
    $\Gamma_b$ & poles & finite & $\Gamma_b(x+b)/\Gamma_b(x)$&${}= \sqrt{2\pi}b^{xb-1/2}/\Gamma(xb)$ \\
    $S_b$ & poles & zeros & $S_b(x+b) / S_b(x)$&${}= 2\sin(\pi bx)$ \\
    $\Upsilon_b$ & zeros & zeros & $\Upsilon_b(x+b) / \Upsilon_b(x)$&${}= b^{1-2bx} \gamma(bx)$ \\
    \bottomrule
  \end{tabular}
\end{center}

\section{Braiding matrices}
\label{sec:matrices}

We determine here the braiding of Toda CFT\@\footnote{Most results apply to all CFTs with
  $W_N$~symmetry, but to be fully general would require various multiplicity indices:
  there may be several primary operators with the same momentum.}  four-point conformal
blocks involving a semi-degenerate operator and a fully degenerate one labelled by the $K$-th symmetric
representation $\repr(Kh_1)$ of~$A_{N-1}$.  Fusion rules of the fully degenerate
operator\footnote{Fusion rules are typically found using a Coulomb gas representation of
  Toda CFT correlators.  They have only been proven rigorously in representation theory of
  $W_N$~algebras in the simplest case: the fundamental representation,
  for $b^2=-1$ (so as to have a free boson realization)~\cite{Gavrylenko:2018ckn}.} reduce
the four-point function to a finite sum (rather than an integral) of holomorphic times
antiholomorphic conformal blocks.  There are $\dim\repr(Kh_1)=\binom{N+K-1}{K}$ terms.  Braiding is
thus given by square matrices of this size.

We begin in \autoref{ssec:matrices-fundamental} with the fundamental representation, so
$\alpha=-bh_1$.  In that case, conformal blocks are known to be hypergeometric functions.
This was initially obtained by solving the null-vector differential equations for
$\NToda=3$ and writing the natural generalization of these results~\cite{Fateev:2007ab}.
Fuchsian techniques provide a proof for all~$\NToda$ when $c=N-1$, namely $b^2=-1$
\cite[Theorem 4.1]{Gavrylenko:2018ckn} (which appeared after version~1 of this paper).  We outline a mild
variation of their proof that generalizes the result to all~$b$, under the assumption that
well-known fusion rules are correct.  Then, as a preparation for more difficult
calculations, we review the braiding matrix for these hypergeometric functions
that forms the foundation of our new results.

We do not discuss here the case $\alpha=-bh_1-\cdots-bh_K$ of antisymmetric representations,
treated in collaboration with Gomis in~\cite{Gomis:2014eya}.
We conjectured there that conformal blocks are vortex partition functions of 2d
$\Nsusy=(2,2)$ SQCD\@.  In Appendix A.3 of that paper we computed the braiding matrix of
the gauge theory result, then showed using shift relations that it is the correct braiding
matrix for conformal blocks.  This method fixes braiding matrices hence monodromies, but
one would need to adapt the proof from the fundamental case to show that the conformal
blocks conjectured from gauge theory are indeed correct.

In \autoref{ssec:matrices-symmetric} we turn to the $K$-th symmetric representation, so
$\alpha=-Kbh_1$, starting from a conjecture for conformal blocks as gauge theory vortex
partition functions~\cite{Gomis:2014eya}.  The hard technical point done in this section
is to work out the braiding matrix for these blocks.  Later, in \autoref{ssec:kernel-shift},
we prove a pentagon identity relating braiding matrices for $K$ and
$K-1$, which proves by induction that the gauge-theory braiding matrix is the correct one
for $W_N$~conformal blocks.  Again, this approach is not sufficient to prove conjectured
expressions of conformal blocks, only their braiding and monodromy behaviour\footnote{We
  thank Sylvain Ribault for discussions on this topic.  It also seems feasible to use the
  recent development~\cite{Gavrylenko:2018ckn} to prove our conjectured conformal blocks.}.

\subsection{Braiding a fundamental degenerate}
\label{ssec:matrices-fundamental}

We focus for now on
$\vev{\widehat{V}\widehat{V}\widehat{V}\widehat{V}} =
\vev{\widehat{V}_{\alpha_{\infty}}(\infty) \widehat{V}_{(\varkappa+b)h_1}(1)
  \widehat{V}_{-bh_1}(x,\bar{x}) \widehat{V}_{\alpha_{0}}(0)}$,
a four-point correlation function with one semi-degenerate momentum $(\varkappa+b)h_1$
and a degenerate $-bh_1$,
labelled by the highest weight $h_1$ of the fundamental representation of~$A_{\NToda-1}$.
The shift by~$b$ in~$\varkappa$ simplifies some expressions.
Momenta are otherwise generic.

Our discussion relies on two fusion rules~\cite{Fateev:2007ab} which can be obtained for the Toda CFT using the
Coulomb gas formalism:
\begin{align}
  \label{matrices-fund-full-fusion}
  \widehat{V}_{-bh_1} \times \widehat{V}_{\alpha} & = \sum_{i=1}^{\NToda} [\widehat{V}_{\alpha-bh_i}] \\
  \label{matrices-fund-simple-fusion}
  \widehat{V}_{-bh_1} \times \widehat{V}_{(\varkappa+b)h_1} & = [\widehat{V}_{\varkappa h_1}] + [\widehat{V}_{(\varkappa+b)h_1 - bh_2}] \,,
\end{align}
where $[\widehat{V}_{\cdots}]$~denotes contributions from $W_N$~descendants of a primary
operator with the given momentum.  As usual in 2d CFT, the four-point function can be
expanded in three different channels by taking the OPE of $\widehat{V}_{-bh_1}$ with any
of the three other operators.  The fusion rules restrict the internal momentum to
$\alpha_{0}-bh_s$ for $1\leq s\leq\NToda$ in the s-channel, $\alpha_{\infty}-bh_s$ in the
u-channel, and $\varkappa h_1$ or $(\varkappa+b)h_1-bh_2$ in the t-channel.

In the s-channel, the four-point function is a sum of~$N$ factorized terms:
\begin{align}
  \vev{\widehat{V}\widehat{V}\widehat{V}\widehat{V}}
  = \sum_{j=1}^{\NToda} C_j^{\sch} \Fblock{\sch}{j}{}{x} \Fblock{\sch}{j}{}{\bar{x}} \,,
  \qquad
  \Fblock{\sch}{j}{}{x} =
  \Fblock{}{}{
    \mathtikz[x=2em,y=1ex,thick]{
      \draw[->-=.55] (1,0) -- (2,0);
      \draw[->-=.55] (5,0) -- (4,0);
      \draw[->-=.55] (4,0) -- (2,0);
      \draw[->-=.55] (2,4) -- (2,0);
      \draw[->-=.55] (4,4) -- (4,0);
      \node at (1.3,1.5) {$\alpha_{\infty}$};
      \node at (4.7,1.5) {$\alpha_{0}$};
      \node at (3,1.5) {$\alpha_{0}-bh_j$};
      \node at (4.5,5) {$-bh_1$};
      \node at (2,5) {$(\varkappa+b)h_1$};
    }
  }{x} \,, &
  \\
  C_j^{\sch} = \frac{\vev{\widehat{V}_{\alpha_{\infty}} \widehat{V}_{(\varkappa+b)h_1} \widehat{V}_{\alpha_{0}-bh_j}}
    \vev{\widehat{V}_{2Q-\alpha_{0}+bh_j} \widehat{V}_{-bh_1} \widehat{V}_{\alpha_{0}}}}
  {\vev{\widehat{V}_{\alpha_{0}-bh_j} \widehat{V}_{2Q-\alpha_{0}+bh_j}}}
  \,, &
\end{align}
where we have absorbed all of the position dependence in the conformal
blocks~$\Fblock{}{}{}{}$.  A useful property of conformal blocks is their $x\to 0$
expansion
\begin{equation}\label{matrices-fund-expans}
  \begin{aligned}
    \Fblock{\sch}{j}{}{x}
    & = x^{\dimToda(\alpha_0-bh_j)-\dimToda(\alpha_0)-\dimToda(-bh_1)} (1+\cdots) \\
    & = x^{b\vev{\alpha_0-Q,h_j} + \frac{\NToda-1}{2} (b^2+1)} (1+\cdots)
  \end{aligned}
\end{equation}
where $(1+\cdots)$ is a series in non-negative integer powers of~$x$, and similarly for
$\Fblock{\sch}{j}{}{\bar{x}}$.  Because of radial ordering, the functions
$\Fblock{\sch}{j}{}{}$ are a priori only defined on the unit disk (with a branch point
at~$0$), but since $\vev{\widehat{V}\widehat{V}\widehat{V}\widehat{V}}$ is smooth away
from $0$, $1$, and $\infty$ the functions can be analytically continued to any simply
connected domain avoiding these points.  Two natural choices are the complex plane minus
cuts on $(-\infty,0]\cup[1,\infty)$, and the complex plane minus cuts on
$[0,1]\cup[1,\infty)$.  We will mostly use the second one.

The u-channel decomposition is similar, and conformal blocks have a simple $x\to\infty$
expansion in terms of a series $1+\cdots$ with non-negative integer powers of $1/x$:
\begin{gather}
  \vev{\widehat{V}\widehat{V}\widehat{V}\widehat{V}} = \sum_{j=1}^{\NToda} C_j^{\uch} \Fblock{\uch}{j}{}{x} \Fblock{\uch}{j}{}{\bar{x}}
  \\
  \label{matrices-fund-expanu}
  \begin{aligned}
    \Fblock{\uch}{j}{}{x} & = x^{\dimToda(\alpha_{\infty})-\dimToda(\alpha_{\infty}-bh_j)-\dimToda(-bh_1)} (1+\cdots) \\
    & = x^{-b\vev{\alpha_{\infty}-Q,h_j} + \frac{\NToda-1}{2} (b^2+1) + \frac{\NToda-1}{\NToda}b^2} (1+\cdots) \,.
  \end{aligned}
\end{gather}
Again, $\Fblock{\uch}{j}{}{x}$ extends to $\mathbb{C}\setminus\{0,1\}$ minus some cuts, for
instance along $[0,1]\cup[1,\infty)$.

The t-channel is more subtle, as it involves three-point functions of
$\widehat{V}_{\alpha_\infty}$, $\widehat{V}_{\alpha_0}$, and a descendant of the primary
$\widehat{V}_{\varkappa h_1}$ or~$\widehat{V}_{(\varkappa+b)h_1-bh_2}$.  Contributions
from primaries with momentum $\varkappa h_1$ factorize because this momentum is
semidegenerate.  Contributions from primaries with momentum $(\varkappa+b)h_1-bh_2$ do not
factorize (except for $\NToda=2$).  We deduce
\begin{equation}
  \vev{\widehat{V}\widehat{V}\widehat{V}\widehat{V}}
  = C_1^{\tch} \Fblock{\tch}{1}{}{x} \Fblock{\tch}{1}{}{\bar{x}} + C_2^{\tch} \mathbb{F}_2^{\tch}(x,\bar{x})
\end{equation}
with the following $x\to 1$ expansions, where $1+\cdots$ denote series in non-negative
integer powers of $(1-x)$ and $(1-\bar{x})$,
\begin{gather}
  \label{matrices-fund-expant1}
  \begin{aligned}
    \Fblock{\tch}{1}{}{x} \Fblock{\tch}{1}{}{\bar{x}}
    & = \abs{1-x}^{2[\dimToda(\varkappa h_1)-\dimToda((\varkappa+b)h_1)-\dimToda(-bh_1)]} (1+\cdots) \\
    & = \abs{1-x}^{2b(\varkappa+b)(\NToda-1)/\NToda} (1+\cdots)
  \end{aligned}
  \\
  \label{matrices-fund-expant2}
  \begin{aligned}
    \mathbb{F}_2^{\tch}(x,\bar{x})
    & = \abs{1-x}^{2[\dimToda((\varkappa+b)h_1-bh_2)-\dimToda((\varkappa+b)h_1)-\dimToda(-bh_1)]} (1+\cdots) \\
    & = \abs{1-x}^{2[-b(\varkappa+b)/\NToda+b^2+1]} (1+\cdots) \,.
  \end{aligned}
\end{gather}

\subsubsection{Monodromy matrices}

Away from the cuts, the equality
$\sum_j C_j^{\sch} \abs*{\Fblock{\sch}{j}{}{}}^2 = \sum_j C_j^{\uch}
\abs*{\Fblock{\uch}{j}{}{}}^2$ implies that the sets $\bigl\{\Fblock{\uch}{j}{}{}\bigr\}$
and $\bigl\{\Fblock{\sch}{k}{}{}\bigr\}$ are different bases of an $\NToda$-dimensional
space of holomorphic functions.  The change of basis matrix is called the braiding matrix.
Similarly $\mathbb{F}_2^{\tch}$ is (non-canonically) a sum of
$\NToda-1$ factorized terms, and their $N-1$ holomorphic factors together with
$\Fblock{\tch}{1}{}{}$ form another basis of the same space.

The expansions of conformal
blocks at $0$, $1$, and~$\infty$ above imply certain monodromy properties when
analytically continuing the functions through cuts.  The monodromy~$M_{(0)}$ around $x=0$
is diagonal in the basis $\Fblock{\sch}{}{}{}$ and eigenvalues can be read off from the
expansion~\eqref{matrices-fund-expans}.  While the monodromies $M_{(1)}$
and~$M_{(\infty)}$ are non-diagonal $\NToda\times\NToda$ matrices in this s-channel basis,
they are diagonalizable (by going to the t- or u-channel bases) and their eigenvalues
can be read off from the expansions above.  Namely, eigenvalues of the
monodromy~$M_{(1)}$ around $x=1$ are seen from \eqref{matrices-fund-expant1}
and~\eqref{matrices-fund-expant2}, the second one having multiplicity $\NToda-1$, and the
monodromy~$M_{(\infty)}$ has $\NToda$ eigenvalues known from~\eqref{matrices-fund-expanu}.
Finally, $M_{(\infty)} = M_{(1)} M_{(0)}$ since $x\in\{0,1,\infty\}$ are the only singular
points.

We thus have a triplet of invertible diagonalizable $\NToda\times\NToda$ matrices
$M_{(0)}$, $M_{(1)}$ and $M_{(\infty)}$ obeying $M_{(\infty)} = M_{(1)} M_{(0)}$ and with
prescribed eigenvalues, such that $M_{(1)}$~has eigenvalues $y_1$~with multiplicity~$1$
and $y_2$~with multiplicity $\NToda-1$ while eigenvalues of $M_{(0)}$ and $M_{(\infty)}$
have no multiplicity.  We now show that such a triplet is unique up to
conjugation.  Choose a basis where $M_{(0)}$~is diagonal.  Since $M_{(1)}-y_2$ has
rank~$1$ it can be written as $M_{(1)ij} = y_2 \delta_{ij} + (y_1-y_2) v_i w_j$ for some
vectors $v$ and~$w$.  By rescaling the $i$-th basis vector by $\sqrt{v_i/w_i}$ and setting
$u_i=\sqrt{v_i w_i}$ we get a basis in which $M_{(0)}=\diag(x_1,\ldots,x_{\NToda})$ is
still diagonal and
\begin{equation}\label{matrices-fund-M1}
  M_{(1)ij} = y_2 \delta_{ij} + (y_1-y_2) u_i u_j \,.
\end{equation}
Denote eigenvalues of $M_{(\infty)}$ as~$z_i$.  Compute the determinant of
$M_{(1)}-y_2x_p M_{(0)}^{-1}$ for some~$p$.  On the one hand the matrix is the sum of a
diagonal matrix $y_2(1-x_p M_{(0)}^{-1})$ with $p$-th entry~$0$ and a rank~$1$ matrix so
subtracting the $p$-th row then the $p$-th column from others makes it diagonal.
On the other hand the matrix is equal to
$(M_{(\infty)}-y_2x_p)M_{(0)}^{-1}$.  We get
\begin{align}
  & \det(M_{(1)}-y_2 x_p M_{(0)}^{-1})
  = (y_1-y_2) u_p^2 \prod_{s\neq p}^{\NToda} (y_2(1-x_p x_s^{-1})) \\
  & \quad = \det(M_{(\infty)} - y_2 x_p) \det(M_{(0)})^{-1}
  = \prod_{s=1}^{\NToda} \frac{z_s - y_2 x_p}{x_s} \,.
\end{align}
Therefore,
\begin{equation}
  u_p^2 = \frac{ \prod_{s=1}^{\NToda} (z_s - y_2 x_p) } {y_2^{\NToda-1} (y_1-y_2) x_p \prod_{s\neq p}^{\NToda} (x_s-x_p)} \,.
\end{equation}
This fixes components of~$u$ up to signs, which can be absorbed in a choice of basis,
hence it fixes~$M_{(1)}$ and concludes this straightforward proof of uniqueness.

Armed with this set of monodromy matrices we should ask whether they fix conformal blocks
uniquely.  The answer is no, of course, as multiplying conformal blocks by a meromorphic
function does not change their monodromy.  However, we can use more information: the
four-point function should only be singular at $z\in\{0,1,\infty\}$ so conformal blocks
must be holomorphic (with branch cuts) away from these points.

\subsubsection{Explicit conformal blocks (proof)}

The s-channel conformal blocks proposed in~\cite{Fateev:2007ab} are
\begin{equation}\label{matrices-fund-sch}
  \mathcal{F}^{\sch,\text{FL}}_j(x)
  = \begin{aligned}[t]
    & x^{b\vev{\alpha_{0}-Q,h_j}+\frac{\NToda-1}{2}(b^2+1)} (1-x)^{b^2+1-b(\varkappa+b)/\NToda}
    \\
    & \times \Hypergeometric[{}_{\NToda}F_{\NToda-1}]
    {1 - b\varkappa/\NToda + b\vev{\alpha_{0}-Q,h_j} + b\vev{\alpha_{\infty}-Q,h_k}, \, 1\leq k\leq\NToda}
    {1 + b\vev{\alpha_{0}-Q,h_j-h_k}, \, k\neq j}
    {x} \,,
  \end{aligned}
\end{equation}
where the hypergeometric function ${}_{\NToda}F_{\NToda-1}$ is defined in terms of
Pochhammer symbols $(a)_k = \Gamma(a+k)/\Gamma(a)$ by the series
\begin{equation}
  \Hypergeometric{a_1 & \cdots & a_{\NToda}}{b_1 & \cdots & b_{\NToda-1}}{x}
  = \sum_{k\geq 0} \frac{x^k}{k!} \frac{(a_1)_k\cdots (a_{\NToda})_k}{(b_1)_k\cdots(b_{\NToda-1})_k} \,.
\end{equation}
They have the correct leading behaviour~\eqref{matrices-fund-expans} near $x=0$, hence the
correct monodromies $M_{(0)}^{\text{FL}}=M_{(0)}$.  As we review in \autoref{sssec:braiding-fund} their analytic continuations close to
$x=1$ and $x=\infty$ are known~\cite{Norlund:1955hyp} and involve the correct powers of
$1-x$ and of $1/x$, respectively, which means that the monodromy matrices have the correct
eigenvalues.  Given our uniqueness proof above, all three monodromy matrices (expressed
for instance in the s-channel basis) are correct.

Using Fuchsian techniques inspired by~\cite{Gavrylenko:2018ckn}, let us now deduce that
the proposed blocks~\eqref{matrices-fund-sch} are indeed the correct conformal
blocks\footnote{Our argument in version~1 was flawed: treating conformal blocks all
  together as in~\cite{Gavrylenko:2018ckn} is essential.  The only assumption in the proof
  we give now is that the well-known fusion rules \eqref{matrices-fund-full-fusion}
  and~\eqref{matrices-fund-simple-fusion} hold.}.  Hypergeometric functions famously obey
a differential equation, and in fact the $N$~functions $\mathcal{F}^{\sch,\text{FL}}_j$
are linearly-independent solutions to the same $N$-th order Fuchsian differential equation
in~$x$, regular except for singularities at $0$, $1$ and~$\infty$.  Thus, the solutions
have Wronskian $\det\Phi\neq 0$ away from these singularities, where\footnote{Note that
  each column of the analogue of~$\Phi$ in~\cite{Gavrylenko:2018ckn} is a collection of
  blocks for a given value of~$\alpha_\infty$.  We could have done the same, with
  $\alpha_\infty=\alpha-h_m/b$ for the $m$-th column where $\alpha$~is some fixed
  momentum.}
\begin{equation}
  \Phi = \begin{pmatrix}
    \mathcal{F}^{\sch,\text{FL}}_1 & \partial_x \mathcal{F}^{\sch,\text{FL}}_1 & \dots &
    \partial_x^{N-1} \mathcal{F}^{\sch,\text{FL}}_1 \\
    \vdots & \vdots & & \vdots \\
    \mathcal{F}^{\sch,\text{FL}}_N & \partial_x \mathcal{F}^{\sch,\text{FL}}_N & \dots &
    \partial_x^{N-1} \mathcal{F}^{\sch,\text{FL}}_N
  \end{pmatrix} .
\end{equation}
Each of the columns has the same monodromy matrices $M_{(0)}$, $M_{(1)}$, $M_{(\infty)}$
as the real conformal blocks~$\mathcal{F}^{\sch}_j$.  Now consider the vector
$\Phi^{-1}\bigl(\mathcal{F}^{\sch}_1,\dots,\mathcal{F}^{\sch}_N\bigr)^T$.  It is holomorphic
away from $x=0$, $x=1$, $x=\infty$ because $\Phi$ is invertible, and it has no monodromy
around these three points because the monodromy matrices cancel.  Expanding near $x=0$,
the non-integer powers $x^{-\beta_j}$ with
$\beta_j=-b\vev{\alpha_0-Q,h_j}-\frac{N-1}{2}(b^2+1)$ cancel and we find
\begin{equation}\label{vec}
  \begin{aligned}
    & \Phi^{-1}\begin{pmatrix}\mathcal{F}^{\sch}_1\\\vdots\\\mathcal{F}^{\sch}_N\end{pmatrix}
    =
    \begin{pmatrix}
      (1 + O(x)) & \frac{-\beta_1}{x}(1+O(x)) & \dots & \frac{(-\beta_1)_N}{x^{N-1}}(1+O(x)) \\
      \vdots & \vdots & & \vdots \\
      (1 + O(x)) & \frac{-\beta_N}{x}(1+O(x)) & \dots & \frac{(-\beta_N)_N}{x^{N-1}}(1+O(x))
    \end{pmatrix}^{-1}
    \begin{pmatrix} 1 + O(x) \\ \vdots \\ 1 + O(x) \end{pmatrix}
    \\
    & \quad =
    \begin{pmatrix}
      1 & 0 & \cdots & 0 \\
      0 & x & \ddots & \vdots \\
      \vdots & \ddots & \ddots & 0 \\
      0 & \cdots & 0 & x^{N-1}
    \end{pmatrix}
    \left(
      \begin{pmatrix}
        1 & -\beta_1 & \dots & (-\beta_1)_N \\
        \vdots & \vdots & & \vdots \\
        1 & -\beta_N & \dots & (-\beta_N)_N
      \end{pmatrix}^{-1}
      \begin{pmatrix} 1 \\ \vdots \\ 1 \end{pmatrix} + O(x) \right)
    = \begin{pmatrix} 1 \\ 0 \\ \vdots \\ 0 \end{pmatrix} + O(x) .
  \end{aligned}
\end{equation}
The expansion near $1$ (resp.~$\infty$) works similarly: the columns with derivatives
involve different powers of $1-x$ (resp.~$1/x$) but this can be factored, leaving a finite
result that tends to the column vector $(1,0,\dots,0)^T$.

Altogether we have a vector whose components are holomorphic on the whole Riemann sphere, hence are
constant.  We know its value at $x=0$ from~\eqref{vec} and deduce that the true conformal blocks
are equal to the first column of~$\Phi$, namely the expression found
in~\cite{Fateev:2007ab}.

\subsubsection{\label{sssec:braiding-fund}Braiding matrices}

Let~$f^{\sch}_j$ denote the hypergeometric function in~\eqref{matrices-fund-sch}.  We now
derive the braiding matrix for the functions
$(-x)^{b\vev{\alpha_{0}-Q,h_j}} f^{\sch}_j(x)$, then convert it to the braiding
matrix~\eqref{matrices-fund-expr} for the blocks~$\mathcal{F}$ by introducing phases.

To compute their braiding matrix it is convenient to introduce notations.  Let $\I m_p$
and $\I\anti{m}_p$ be $2\NToda$ complex numbers such that
\begin{equation}\label{mass-dictionary}
  \alpha_{0} = Q - \frac{1}{b} \sum_{p=1}^{\NToda} \I m_p h_p , \quad
  \alpha_{\infty} = Q - \frac{1}{b} \sum_{p=1}^{\NToda} \I\anti{m}_p h_p , \quad
  \varkappa = \frac{1}{b} \sum_{p=1}^{\NToda} (1 + \I m_p + \I\anti{m}_p) .
\end{equation}
This parametrization is
redundant under shifts of all $\I m_p$ and $-\I\anti{m}_p$.  The s-channel factor
$(-x)^{-\I m_j} f^{\sch}_j(x)$ can be expressed as the Mellin--Barnes integral given
below, which converges away from the positive real axis.  For $\abs{x}\lessgtr 1$ we can
close the contour integral towards $\upsilon\to\pm\infty$, and enclose either the poles at
$\upsilon+\I m_j \in \bbZ_{\geq 0}$ or the $\NToda$~families of poles at
$\upsilon-\I\anti{m}_k \in \bbZ_{\leq 0}$ labelled by $1\leq k\leq\NToda$.  The first choice
yields a single s-channel factor, while the second yields a sum of $\NToda$ u-channel
factors:
\begin{align}
  (-x)^{-\I m_j} f^{\sch}_p(x)
  & \stackrel{\text{cont}}{=} D_j
  \int_{-\I\infty}^{\I\infty} \frac{\dd{\upsilon}}{2\pi\I}
  \frac{\prod_{k=1}^{\NToda}\Gamma(-\I\anti{m}_k+\upsilon)}
    {\prod_{k\neq j}^{\NToda}\Gamma(1+\I m_k+\upsilon)}
  \Gamma(-\upsilon-\I m_j) (-x)^{\upsilon}
  \\\label{matrices-fund-MB-lhs}
  & \stackrel{\text{cont}}{=}
  \sum_{k=1}^{\NToda} D_j \check{B}_{jk}^0 \anti{D}_k
  (-x)^{\I\anti{m}_k} f^{\uch}_k(x) \,.
\end{align}
We will not need the explicit expression for~$f^{\uch}_k$, which are hypergeometric series $1+\cdots$ in
non-positive integer powers of~$x$.  The coefficients $D$, $\check{B}^0$ and~$\anti{D}$
are given by
\begin{equation}
\begin{gathered}
  \check{B}^\epsilon_{jk} =
  \frac{\pi e^{\pi\epsilon(m_j+\anti{m}_k)}}{\sin\pi(-\I\anti{m}_k-\I m_j)}
  \\
  \begin{aligned}
    D_j & = \prod_{t=1}^{\NToda}
    \frac{\Gamma(1+\I m_t-\I m_j)}{\Gamma(-\I\anti{m}_t-\I m_j)}
    &
    \anti{D}_k & =
    \frac{\prod_{t\neq k}^{\NToda} \Gamma(-\I\anti{m}_t+\I\anti{m}_k)}
    {\prod_{t=1}^{\NToda} \Gamma(1+\I m_t+\I\anti{m}_k)} \,.
  \end{aligned}
\end{gathered}
\end{equation}
Here we have included a parameter $\epsilon\in\{0,\pm 1\}$.  It is sometimes convenient to
consider s-channel factors $x^{-\I m_j} f^{\sch}_j(x)$ analytically continued with branch
cuts on $(-\infty,0] \cup [1,+\infty)$, and u-channel factors
$x^{\I\anti{m}_k} f^{\uch}_k(x)$ with branch cuts along $(-\infty,0] \cup [0,1]$.  Using
$(-x)^{\lambda} = e^{-\I\pi\epsilon\lambda} x^{\lambda}$ for $\epsilon=\sign(\Im x)$, we
obtain
\begin{equation}\label{matrices-fund-feps}
  x^{-\I m_j} f^{\sch}_j(x) \stackrel{\text{cont}}{=} \sum_{k=1}^{\NToda} D_j \check{B}^\epsilon_{jk} \anti{D}_k x^{\I\anti{m}_k} f^{\uch}_k(x) \,.
\end{equation}
The braiding matrix in $\mathcal{F}^{\sch}_{j}(x)= \sum_{k=1}^{\NToda} \Braiding{\epsilon}{jk}{} \mathcal{F}^{\uch}_k(x)$ includes a phase: explicitly we get
\begin{equation}
  \label{matrices-fund-expr}
  \begin{aligned}
    \Braiding{\epsilon}{jk}{(\varkappa+b)h_1&-bh_1\\\alpha_{\infty}&\alpha_{0}}
    & = e^{\I\pi\epsilon[b(\varkappa+b)/\NToda - b^2 - 1]} D_j \check{B}^\epsilon_{jk} \anti{D}_k \\
    & = \begin{aligned}[t]
      &
      \frac{\prod_{t\neq j}^{\NToda} \Gamma(1 + b\vev{Q-\alpha_{0},h_t-h_j})}
      {\prod_{u=1}^{\NToda} \Gamma(1 - \frac{b\varkappa}{\NToda}-b\vev{Q-\alpha_{0},h_j}-b\vev{Q-\alpha_{\infty},h_u})}
      \\
      & \times
      \frac{\pi e^{\I\pi\epsilon [-b^2(\NToda-1)/\NToda-b\vev{Q-\alpha_{0},h_j}-b\vev{Q-\alpha_{\infty},h_k}]}}
      {\sin\pi(1-\frac{b\varkappa}{\NToda}-b\vev{Q-\alpha_{0},h_j}-b\vev{Q-\alpha_{\infty},h_k})}
      \\
      & \times
      \frac{\prod_{u\neq k}^{\NToda} \Gamma(b\vev{Q-\alpha_{\infty},h_k-h_u})}
      {\prod_{t=1}^{\NToda} \Gamma(\frac{b\varkappa}{\NToda}+b\vev{Q-\alpha_{0},h_t}+b\vev{Q-\alpha_{\infty},h_k})} \,.
    \end{aligned}
  \end{aligned}
\end{equation}

The s-channel blocks $\mathcal{F}^{\sch}_j(x)$ have the expected power of~$x$ (hence the
expected monodromies) around~$0$.  The eigenvalues of the monodromy~$M_{(\infty)}$
around~$\infty$ are also correct, as can be checked by combining the u-channel factors
in~\eqref{matrices-fund-MB-lhs} with the additional powers of $x$ and $1-x$
in~\eqref{matrices-fund-sch} and comparing to the u-channel
asymptotics~\eqref{matrices-fund-expanu} expected from CFT\@.  Finally, from the braiding
matrix above we deduce the monodromy $M_{(1)}=\Braiding{+}{}{}(\Braiding{-}{}{})^{-1}$
around~$1$.  All components $\check{B}^+_{jk} - \check{B}^-_{jk} =2\pi\I$ are equal hence
$\check{B}^+-\check{B}^-$ has rank~$1$, and
$e^{-2\pi\I\gamma_1}M_{(1)}-1=D(\check{B}^+-\check{B}^-)\anti{D}(\Braiding{-}{}{})^{-1}$
too.  Therefore, $M_{(1)}$~has the eigenvalue~$e^{2\pi\I\gamma_1}$ with multiplicity
$\NToda-1$.  Its last eigenvalue is fixed by
$\det e^{-2\pi\I\gamma_1}M_{(1)} = \det \check{B}^+ / \det \check{B}^-$.  These
eigenvalues coincide with those expected of t-channel blocks.  In fact the precise
exponents of $(1-x)$ also match~\cite{Norlund:1955hyp}.  We used this in our proof that
conformal blocks are given by the explicit hypergeometric
formula~\eqref{matrices-fund-sch}.

\subsection{Braiding a symmetric degenerate}
\label{ssec:matrices-symmetric}

We now generalize the discussion above to a degenerate vertex
operator~$\widehat{V}_{-Kbh_1}$ labelled by the $K$-th symmetric representation
$\repr(Kh_1)$ of~$A_{\NToda-1}$.  Namely we consider its four-point function with two
generic operators $\widehat{V}_{\alpha_{\infty}}$ and~$\widehat{V}_{\alpha_{0}}$ and one
semi-degenerate~$\widehat{V}_{(\varkappa+Kb)h_1}$, including a shift by~$Kb$ for later
convenience.  The braiding~\eqref{matrices-sym-expr} we find is new.  Contrarily to
the $K=1$ case of the previous section which we reuse throughout the paper, the
$K>1$ case is only used to find the correct generalization to continuous values
of~$K$ in \autoref{sec:kernel}.

Explicit s-channel and u-channel decompositions and their conformal blocks were
conjectured in \cite[Appendix A.5]{Gomis:2014eya}, thanks to a relation with sphere
partition functions of 2d $\Nsusy=(2,2)$ gauge theories.  We use the same notations $\I m_p$
and $\I\anti{m}_p$ as in~\eqref{mass-dictionary} (ambiguous under
overall shifts of all $\I m_p$ and $-\I\anti{m}_p$), namely
$\alpha_{0} = Q - \frac{1}{b} \sum_{p=1}^{\NToda} \I m_p h_p$ and
$\alpha_{\infty} = Q - \frac{1}{b} \sum_{p=1}^{\NToda} \I\anti{m}_p h_p$ and
$\varkappa = \frac{1}{b} \sum_{p=1}^{\NToda} (1 + \I m_p + \I\anti{m}_p)$.
The four-point function is a sum over weights $h_{[n]} = \sum_{s=1}^{\NToda} n_s h_s$ of
the symmetric representation $\repr(Kh_1)$: up to a constant~$C$,
\begin{equation}\label{matrices-sym-VVVV}
  \begin{aligned}
    & \vev{\widehat{V}_{\alpha_{\infty}}(\infty) \widehat{V}_{(\varkappa+Kb)h_1}(1) \widehat{V}_{-Kbh_1}(x,\bar{x}) \widehat{V}_{\alpha_{0}}(0)}
    \\
    & \quad = C \sum_{n_1+\cdots+n_{\NToda}=K} \biggl[
    \prod_{(s,\mu)}^{[n]} \prod_{t=1}^{\NToda} \frac{\gamma(\I m_{s\mu}-\I m_{tn_t})} {\gamma(1+\I\anti{m}_t+\I m_{s\mu})}
    \Fblock{\sch}{[n]}{}{x} \Fblock{\sch}{[n]}{}{\bar{x}}
    \biggr] \,,
  \end{aligned}
\end{equation}
where $\gamma(y)=\Gamma(y)/\Gamma(1-y)$ and we introduced the notations
\begin{equation}\label{smu-notation}
  \prod_{(s,\mu)}^{[n]} = \prod_{s=1}^{\NToda} \prod_{\mu=0}^{n_s-1} \quad\text{and}\quad
  \I m_{s\mu} = \I m_s + \mu b^2.
\end{equation}
The conformal blocks $\Fblock{\sch}{[n]}{}{}$ are
\begin{align}
  \Fblock{\sch}{[n]}{}{x} & = (1-x)^{-\gamma_1} x^{-\gamma_0-\sum_{(s,\mu)}^{[n]} \I m_{s\mu}} f^{\sch}_{[n]}(x) \\
  \gamma_0 & = -\frac{K(K-1)}{2} b^2 - \frac{K(\NToda-1)}{2} (b^2+1) - \frac{K}{\NToda} \sum_{s=1}^{\NToda} \I m_s \\
  \gamma_1 & = - \frac{K(\NToda-K)}{\NToda} b^2 + \frac{K}{\NToda} \sum_{s=1}^{\NToda}(\I m_s+\I\anti{m}_s)
  \displaybreak[0]\\
  \nonumber
  f^{\sch}_{[n]}(x)
  & = \sum_{k_{\cdots}\geq 0} \prod_{(s,\mu)}^{[n]} \biggl[ x^{k_{s\mu}}
    \prod_{t=1}^{\NToda} \frac{(-\I\anti{m}_t-\I m_{s\mu})_{k_{s\mu}}}{(1+\I m_{tn_t}-\I m_{s\mu})_{k_{s\mu}}}
  \\\label{matrices-sym-fsch}
  & \qquad \qquad \quad \times
    \frac{\prod_{t=1}^{\NToda} (1+\I m_{tn_t}-\I m_{s\mu}+k_{s\mu}-k_{t(n_t-1)})_{k_{t(n_t-1)}}}
    {\prod_{(t,\nu)}^{[n]} (1+\I m_{t\nu}-\I m_{s\mu}+k_{s\mu}-k_{t\nu})_{k_{t\nu}-k_{t(\nu-1)}}}
    \biggr] \,.
\end{align}

For a given weight $h_{[n]}$ of $\repr(Kh_1)$, and a choice of $1\leq p\leq\NToda$ we
consider the following generalization of the Mellin--Barnes
integral~\eqref{matrices-fund-MB-lhs} used for $K=1$.
\begin{equation}
  \begin{aligned}
    I^p_{[n]}
    & = \frac{\Gamma(-b^2)^K}{K!}
    \prod_{j=1}^K \biggl[ \int_{-\infty}^{\infty} \frac{\dd{\sigma_j}}{2\pi} \biggr] \Biggl\{
    \prod_{i\neq j}^K \frac{\Gamma(\I\sigma_i-\I\sigma_j-b^2)}{\Gamma(\I\sigma_i-\I\sigma_j)}
    \\
    & \quad \times \prod_{j=1}^K \biggl[ (-x)^{\I\sigma_j}
    \frac{\prod_{s=1}^{\NToda} \bigl[\Gamma(-\I\anti{m}_s+\I\sigma_j) \Gamma(-\I m_s-\I\sigma_j) \bigr]}
    {\prod_{s\neq p}^{\NToda} \bigl[\Gamma(1+\I m_{sn_s}+\I\sigma_j) \Gamma(-\I m_{sn_s}-\I\sigma_j)\bigr]}
    \biggr] \Biggr\} \,.
  \end{aligned}
\end{equation}
The contours lie between poles of all the $\Gamma(-\I\anti{m}_s+\I\sigma_j)$ and poles of
all the $\Gamma(-\I m_s-\I\sigma_j)$.  The integral converges for $x\not\in[0,\infty)$
because there are fewer $\Gamma$~functions in the denominator.

For $\abs{x}\lessgtr 1$ we can close contours towards $\I\sigma_j\to\mp\I\infty$,
enclosing some poles.  Start with $\abs{x}<1$.  If the partition $[n]$ has a single
non-zero entry (e.g.\@, if $K=1$) we can choose~$p$ so that $n_s=0$ for $s\neq p$, and
then all numerators $\Gamma(-im_s-i\sigma_j)$ cancel with denominators except for $s=p$.
But in general there is no such cancellation and the integral gives a linear combination
of s-channel factors~\eqref{matrices-sym-fsch}; our whole work will be to disentangle this
sum:
\begin{gather}
  I^p_{[n]} = \sum_{[k]} T^p_{[n][k]} (-x)^{-\sum_{(s,\mu)}^{[k]} \I m_{s\mu}} f^{\sch}_{[k]} , \\
  T^p_{[n][k]} = \prod_{(s,\mu)}^{[k]}
  \frac{\prod_{t=1}^{\NToda} \bigl[ \Gamma(-\I\anti{m}_t-\I m_{s\mu}) \Gamma(\I m_{s\mu} - \I m_{tk_t}) \bigr]}
  {\prod_{t\neq p}^{\NToda} \bigl[ \Gamma(1+\I m_{tn_t} - \I m_{s\mu}) \Gamma(\I m_{s\mu} - \I m_{tn_t}) \bigr]} \,.
\end{gather}
The sum ranges over weights of~$\repr(Kh_1)$, but the matrix~$T^p$ is ``triangular'' in
the sense that its component $T^p_{[n][k]}$ vanishes if $n_s<k_s$ for any $s\neq p$.

Closing contours towards $\I\sigma_j\to\I\infty$ for $\abs{x}>1$ gives a sum of u-channel factors
\begin{gather}
  I^p_{[n]} = \sum_{[\anti{n}]} U^p_{[n][\anti{n}]} (-x)^{\sum_{(s,\mu)}^{[\anti{n}]} \I\anti{m}_{s\mu}} f^{\uch}_{[\anti{n}]} \\
  U^p_{[n][\anti{n}]} = \prod_{(s,\mu)}^{[\anti{n}]}
  \frac{\prod_{t=1}^{\NToda} \bigl[ \Gamma(\I\anti{m}_{s\mu}-\I\anti{m}_{t\anti{n}_t}) \Gamma(-\I\anti{m}_{s\mu}-\I m_t) \bigr]}
  {\prod_{t\neq p}^{\NToda} \bigl[ \Gamma(1+\I m_{tn_t}+\I\anti{m}_{s\mu}) \Gamma(-\I\anti{m}_{s\mu}-\I m_{tn_t}) \bigr]} \,.
\end{gather}
This leads to the braiding
\begin{equation}\label{matrices-sym-sf}
  (-x)^{-\sum_{(s,\mu)}^{[n]} \I m_{s\mu}} f^{\sch}_{[n]}
  = \sum_{[\anti{n}]} \bigl((T^p)^{-1} U^p\bigr)_{[n][\anti{n}]} (-x)^{\sum_{(s,\mu)}^{[\anti{n}]} \I\anti{m}_{s\mu}} f^{\uch}_{[\anti{n}]} \,.
\end{equation}
We thus need to invert the matrix~$T^p$ then multiply the result by~$U^p$.  A consistency
check is that the braiding matrix $(T^p)^{-1} U^p$ must not depend on~$p$.

Split $T^p = \check{T}^p D^p$ with $D^p$~diagonal:
\begin{align}
  \check{T}^p_{[n][k]}
  & = \prod_{(s,\mu)}^{[k]} \prod_{t\neq p}^{\NToda} \frac{1}{\pi} \sin\pi(\I m_{s\mu} - \I m_{tn_t})
  \\
  D^p_{[k][l]}
  & = \delta_{[k][l]} \prod_{(s,\mu)}^{[k]} \prod_{t=1}^{\NToda} \bigl[ \Gamma(-\I\anti{m}_t-\I m_{s\mu}) \Gamma(\I m_{s\mu}-\I m_{tk_t}) \bigr] \,.
\end{align}
A proposal for $(\check{T}^p)^{-1}$ is found by trial and error:
\begin{equation}\label{matrices-sym-Tchinv}
  \bigl((\check{T}^p)^{-1}\bigr)_{[n][k]}
  =
  \begin{cases}
    0 \quad \text{if $n_t<k_t$ for any $t\neq p$, and otherwise}\\
    \frac{\prod_{s<t}^{\NToda} \bigl[\frac{1}{\pi}\sin\pi(\I m_{sk_s}-\I m_{tk_t}) \frac{1}{\pi}\sin\pi(\I m_{tn_t} - \I m_{sn_s})\bigr]}
    {\prod_{t\neq p}^{\NToda} \Bigl[\frac{1}{\pi}\sin\pi(\I m_{tk_t}-\I m_{pk_p})
      \prod_{\substack{\scriptscriptstyle 1\leq s\leq\NToda,0\leq\mu\leq n_s\\\scriptscriptstyle (s,\mu)\neq(t,k_t)}} \frac{1}{\pi} \sin\pi(\I m_{s\mu}-\I m_{tk_t}) \Bigr]}
    \,.
  \end{cases}
\end{equation}
We must prove that
$\sum_{[k]} \bigl((\check{T}^p)^{-1}\bigr)_{[n][k]} (\check{T}^p)_{[k][j]}
=\delta_{[n][j]}$.
Since both matrices are ``triangular'' their product is as well.  It is straightforward to
compute the diagonal coefficients
$\bigl((\check{T}^p)^{-1} \check{T}^p\bigr)_{[n][n]} =
\bigl((\check{T}^p)^{-1}\bigr)_{[n][n]} (\check{T}^p)_{[n][n]} = 1$.
There remains to show that coefficients $[n][j]$ with $j_s\leq n_s$ for
all~$s\neq p$, and with $[j]\neq [n]$ (equivalently $j_p>n_p$) vanish.  Cancelling
factors of $\frac{1}{\pi}\sin\pi(\dots)$ as much as possible yields
\begin{align}
  & \sum_{[k]} \bigl((\check{T}^p)^{-1}\bigr)_{[n][k]} (\check{T}^p)_{[k][j]}
  \\\nonumber
  & = \sum_{\substack{[k]\\j_s\leq k_s\leq n_s\,\forall s\neq p}}
    \begin{aligned}[t]
      &
      \prod_{\substack{1\leq s<t\leq\NToda\\s,t\neq p}} \biggl[
      \frac{1}{\pi} \sin\pi(\I m_{sk_s}-\I m_{tk_t}) \frac{1}{\pi}\sin\pi(\I m_{tn_t} - \I m_{sn_s}) \biggr]
      \\
      & \times \prod_{t\neq p}^{\NToda}
      \frac{\frac{1}{\pi}\sin\pi(\I m_{pn_p} - \I m_{tn_t}) \prod_{\mu=n_p+1}^{j_p-1} \frac{1}{\pi} \sin\pi(\I m_{p\mu} - \I m_{tk_t})}
      {\prod_{\substack{s\neq p,j_s\leq \mu\leq n_s\\(s,\mu)\neq (t,k_t)}} \frac{1}{\pi} \sin\pi(\I m_{s\mu}-\I m_{tk_t})} \,.
    \end{aligned}
\end{align}
This is the sum of residues of
\begin{equation}
  \begin{aligned}
    & \prod_{\substack{s<t\\s,t\neq p}}^{\NToda} \biggl[ \frac{1}{\pi} \sin\pi(\I\tau_t-\I\tau_s) \frac{1}{\pi}\sin\pi(\I m_{tn_t} - \I m_{sn_s}) \biggr]
    \\
    & \times \prod_{t\neq p}^{\NToda}
    \frac{\frac{1}{\pi}\sin\pi(\I m_{pn_p} - \I m_{tn_t}) \prod_{\mu=n_p+1}^{j_p-1} \frac{1}{\pi} \sin\pi(\I m_{p\mu}+\I\tau_t)}
    {\prod_{s\neq p}^{\NToda} \prod_{\mu=j_s}^{n_s} \frac{1}{\pi} \sin\pi(\I m_{s\mu}+\I\tau_t)}
  \end{aligned}
\end{equation}
at $\I\tau_t = -\I m_{tk_t}$.  Each $\I\tau_t$ appears in $\NToda-2+j_p-n_p-1$ sines in
the numerator, and $\sum_{s\neq p} (1+n_s-j_s) = \NToda-1+j_p-n_p$ in the denominator, in
other words, two more.  Thus the function is $1$-periodic in each variable~$\I\tau_t$, and
decays exponentially as $\I\tau_t\to\pm\infty$.  The sum of residues thus vanishes,
because it is the sum of all residues in a fundamental domain of the periodicity, and
there is no contribution from infinity.  This establishes~\eqref{matrices-sym-Tchinv}.

The braiding matrix~\eqref{matrices-sym-sf} is then
$B = (T^p)^{-1} U^p = (D^p)^{-1} (\check{T}^p)^{-1} U^p$.  The result is a sum
of residues of some function of $\NToda-1$ variables~$\tau_t$ for $t\neq p$.  Relabelling
the variables~$\tau_t$ using a permutation of $\intset{1}{\NToda}$ so that they are
numbered from $1$ to $\NToda-1$ and $\phi(\NToda)=p$, we obtain
(recall the notations \eqref{mass-dictionary} and~\eqref{smu-notation})
\begin{equation}\label{matrices-sym-expr}
  \begin{aligned}
    B^\phi_{[n][\anti{n}]}
    & = (-1)^{\phi}
    \prod_{t=1}^{\NToda}
    \frac{\prod_{(s,\mu)}^{[\anti{n}]} \Gamma(\I\anti{m}_{s\mu}-\I\anti{m}_{t\anti{n}_t}) \Gamma(-\I\anti{m}_{s\mu}-\I m_t)}
    {\prod_{(s,\mu)}^{[n]} \Gamma(-\I\anti{m}_t-\I m_{s\mu}) \Gamma(\I m_{s\mu}-\I m_{tn_t})}
    \prod_{s<t}^{\NToda} \frac{\sin\pi(\I m_{tn_t} - \I m_{sn_s})}{\pi}
    \!
    \\
    & \qquad \times
    \prod_{j=1}^{\NToda-1} \Biggl[
    \sum_{k_j=0}^{n_{\phi(j)}}\res_{\I\tau_j=-\I m_{\phi(j)k_j}}
    \frac{\prod_{(s,\mu)}^{[\anti{n}]} \frac{1}{\pi}\sin\pi(-\I\anti{m}_{s\mu}+\I\tau_j)}
    { \prod_{s=1}^{\NToda} \prod_{\mu=0}^{n_s} \frac{1}{\pi}\sin\pi(\I m_{s\mu}+\I\tau_j) }
    \prod_{i=1}^{j-1} \frac{\sin\pi(\I\tau_j-\I\tau_i)}{\pi}
    \Biggr]
  \end{aligned}
\end{equation}
where $(-1)^{\phi}$ is the signature of~$\phi$.  This expression does not change if we
replace $\phi$ by another permutation such that $\phi(\NToda)=p$ and we permute the
$\tau_j$ accordingly: indeed, the sign coming from $\sin\pi(\I\tau_j-\I\tau_i)$ is compensated
by the change in $(-1)^{\phi}$.

Let us show that $B^\phi$ does not depend on the arbitrary choice of~$p$ either, hence is
independent of~$\phi$.  Choose an index $1\leq j\leq\NToda-1$.  The variable~$\tau_j$
appears in $\NToda-2+K$ sines in the numerator and $\NToda+K$ sines in the denominator
of~\eqref{matrices-sym-expr}.  We thus have $\I\tau_j\to\I\tau_j+1$ periodicity and no
residue at infinity, hence the sum of residues at $\I\tau_j = -\I m_{\phi(j)k_j}$ is equal
to minus the sum of all other residues in a strip of width~$1$.  This yields a sum over
$\I\tau_j = -\I m_{\phi(i)k}$ for all $1\leq i\leq\NToda$ with $i\neq j$ and
$0\leq k\leq n_{\phi(i)}$.  The contribution from a given~$i$ with $i<\NToda$ (and
$i\neq j$) vanishes by antisymmetry under the exchange $\tau_i\leftrightarrow\tau_j$, thus
only the poles at $-\I m_{\phi(\NToda)k} = -\I m_{pk}$ contribute.  All in all, we obtain
the same expression as~\eqref{matrices-sym-expr}, with $\phi(j)$ and $\phi(\NToda)$
exchanged.  The sign coming from flipping the contour is absorbed into a change of the
signature $(-1)^{\phi}$.

As for $K=1$, the braiding matrix for $\Fblock{\sch}{[n]}{}{x}$ is obtained by including a
phase $e^{\I\pi\epsilon\gamma_1}$, and another phase comes from using factors $x^{\cdots}$
instead of $(-x)^{\cdots}$.  Putting everything together yields
\begin{gather}
  \Fblock{\sch}{[n]}{}{x}
  = \sum_{[\anti{n}]} \Braiding{\epsilon}{[n][\anti{n}]}{} \Fblock{\uch}{[\anti{n}]}{}{x}
  \\\label{matrices-sym-B}
  \Braiding{\epsilon}{[n][\anti{n}]}{}
  = e^{\I\pi\epsilon\gamma_1} e^{\I\pi\epsilon\sum_{(s,\mu)}^{[n]}(-\I m_{s,\mu})}
  B^\phi_{[n][\anti{n}]}
  e^{-\I\pi\epsilon\sum_{(s,\mu)}^{[\anti{n}]}\I\anti{m}_{s,\mu}}
  \,.
\end{gather}
The explicit expression of~$B^\phi$ involves a permutation~$\phi$, but is independent of
it.  To translate this expression explicitly back from the $\{\I m,\I\anti{m}\}$
notation~\eqref{mass-dictionary} to momenta, replace
\begin{equation}
  \begin{aligned}
    \I m_{s\mu} & = b\vev{Q-\alpha_{0}, h_s} + \mu b^2 + \frac{1}{\NToda} \sum_{t=1}^{\NToda} \I m_t \\
    \I\anti{m}_{s\mu} & = \frac{b\varkappa}{\NToda} + b\vev{Q-\alpha_{\infty},h_s} + \mu b^2 -
    1 - \frac{1}{\NToda} \sum_{t=1}^{\NToda} \I m_t
  \end{aligned}
\end{equation}
then shift the variables $\I\tau_j$ to absorb $\frac{1}{\NToda} \sum_{t=1}^{\NToda} \I m_t$.

Note that the starting point of this calculation, namely the explicit
expression~\eqref{matrices-sym-VVVV} for the four-point function, is not proven.  However,
we prove a shift relation in \autoref{sec:kernel} for a generalization of the
braiding~\eqref{matrices-sym-expr}.  This relation expresses in particular the braiding for a
given~$\widehat{V}_{-Kbh_1}$ in terms of that for $\widehat{V}_{-(K-1)bh_1}$ and the
braiding and fusion for~$\widehat{V}_{-bh_1}$.  By induction on~$K$ this proves that the
braiding matrix given here is correct.

\section{Braiding kernel}
\label{sec:kernel}

\autoref{ssec:kernel-formula} conjectures the braiding kernel~\eqref{kernel-expression} of two semi-degenerate
vertex operators, which generalizes the braiding/fusion kernel for Virasoro ($\NToda=2$)
conformal blocks~\cite{Ponsot:1999uf,Ponsot:2000mt}.
We explain in \autoref{ssec:kernel-degenerate}
how it reduces to the braiding matrix~\eqref{matrices-sym-expr} and in
\autoref{ssec:kernel-shift} that it obeys a very constraining shift relation deduced
from a Moore--Seiberg pentagon identity.  Discrete symmetries are investigated later in
\autoref{ssec:wall-duality}.

\subsection{\label{ssec:kernel-formula}Main formula}

The four-point function
with two generic momenta $\alpha_1$, $\alpha_3$ and two semi-degenerate momenta
$\alpha_2=\kappa_2 h_1$, $\alpha_4=\kappa_4 h_1$ has an s-channel decomposition
\begin{equation}\label{kernel-sch}
  \begin{aligned}
    & \vev{\widehat{V}_{\alpha_3}(\infty) \widehat{V}_{\kappa_4 h_1}(1) \widehat{V}_{\kappa_2 h_1}(x,\bar{x}) \widehat{V}_{\alpha_1}(0)} \\[-10pt]
    & = \int \dd{\alpha_{12}} \frac{\widehat{C}(\alpha_3,\kappa_4 h_1,2Q-\alpha_{12}) \widehat{C}(\alpha_{12},\kappa_2 h_1, \alpha_1)}{\vev{\widehat{V}_{2Q-\alpha_{12}}\widehat{V}_{\alpha_{12}}}}
    \abs*{
    \Fblock{}{}
      {
        \mathtikz[x=1em,y=1ex,thick]{
          \draw[->-=.55] (0,0) -- (2,0);
          \draw[->-=.55] (6,0) -- (4,0);
          \draw[->-=.55] (4,0) -- (2,0);
          \draw[->-=.55] (2,-4) -- (2,0);
          \draw[->-=.55] (4,4) -- (4,0);
          \node at (0.3,1.5) {$\alpha_3$};
          \node at (5.7,-1.5) {$\alpha_1$};
          \node at (3,1.5) {$\alpha_{12}$};
          \node at (5.3,3) {$\kappa_2 h_1$};
          \node at (0.7,-3) {$\kappa_4 h_1$};
        }
      }{x}
    }^2
  \end{aligned}
\end{equation}
where the denominator is non-zero because of our Weyl-invariant choice of normalization of
vertex operators, and $|\cdots|^2$ involves conjugating~$x$ but not momenta.  Note that
the internal momentum
$\alpha_{12}$ is continuous rather than discrete because there is no fully degenerate
vertex operator.  The s-channel conformal blocks are in principle fixed by $W_{\NToda}$
symmetry.  In practice, closed forms are only known thanks to the AGT relation with
instanton partition functions, and we will not need them.  In this section we again
normalize conformal blocks as
$\Fblock{\sch}{\alpha_{12}}{}{x} =
(-x)^{\dimToda(\alpha_{12})-\dimToda(\alpha_{1})-\dimToda(\kappa_2 h_1)} (1 + \cdots)$:
the use of $-x$ instead of~$x$ avoids phases.

The u-channel counterpart of~\eqref{kernel-sch} has $\kappa_2\leftrightarrow\kappa_4$:
\begin{equation}
  \begin{aligned}
    & \vev{\widehat{V}_{\alpha_3}(\infty) \widehat{V}_{\kappa_4 h_1}(1) \widehat{V}_{\kappa_2 h_1}(x,\bar{x}) \widehat{V}_{\alpha_1}(0)} \\[-10pt]
    & = \int \dd{\alpha_{32}} \frac{\widehat{C}(\alpha_3,\kappa_2 h_1,\alpha_{32}) \widehat{C}(2Q-\alpha_{32},\kappa_4 h_1, \alpha_1)}{\vev{\widehat{V}_{2Q-\alpha_{32}}\widehat{V}_{\alpha_{32}}}}
    \abs*{
    \Fblock{}{}
      {
        \mathtikz[x=1em,y=1ex,thick]{
          \draw[->-=.55] (0,0) -- (2,0);
          \draw[->-=.55] (6,0) -- (4,0);
          \draw[->-=.55] (2,0) -- (4,0);
          \draw[->-=.55] (2,4) -- (2,0);
          \draw[->-=.55] (4,-4) -- (4,0);
          \node at (0.3,-1.5) {$\alpha_3$};
          \node at (5.7,1.5) {$\alpha_1$};
          \node at (3,-1.5) {$\alpha_{32}$};
          \node at (5.3,-3) {$\kappa_4 h_1$};
          \node at (0.7,3) {$\kappa_2 h_1$};
        }
      }{x}
    }^2 \,.
  \end{aligned}
\end{equation}
Again, we normalize these u-channel conformal blocks so that their leading term is a power
of~$(-x)$, namely
$\Fblock{\uch}{\alpha_{32}}{}{x} \sim
(-x)^{\dimToda(\alpha_3)-\dimToda(\alpha_{32})-\dimToda(\kappa_2 h_1)}$.
Both sets of conformal blocks are analytic on $\bbC\setminus[0,\infty)$.  The two
decompositions are related by an integral transformation
\begin{equation}
  \Fblock{\sch}{\alpha_{12}}{}{x}
  = \int \dd{\alpha_{32}} \Braiding{}{\alpha_{12}\alpha_{32}}{} \Fblock{\uch}{\alpha_{32}}{}{x} \,.
\end{equation}
Our goal is to find the braiding kernel~$\Braiding{}{\alpha_{12}\alpha_{32}}{}$.

From \autoref{ssec:matrices-symmetric} we know this braiding kernel in the limit
$\kappa_2 h_1\to -Kbh_1$, in other words when one of the semi-degenerate operators
turns into a degenerate operator.  Then it is a sum of residues (hence an integral) of a
product of sines~\eqref{matrices-sym-expr} which involve various multiples of~$b^2$ in
their arguments.  This product of sines can be recast in terms of the double Sine
function~$S_b$ which obeys
$S_b(x+nb)/S_b(x) = \prod_{\mu=0}^{n-1} 2\sin\pi (bx+\mu b^2)$ (see \autoref{ssec:functions} on special functions).
The braiding kernel for generic~$\kappa_2$ should thus be an integral of some
$S_b$~functions.  Writing all generic momenta as $\alpha = Q - \I a$, we propose
\begin{equation}\label{kernel-expression}
  \begin{aligned}
    & \Braiding{}{\alpha_{12}\alpha_{32}}{\kappa_4 h_1 & \kappa_2 h_1 \\ \alpha_3 & \alpha_1}
    = \I^{\NToda-1} \prod_{s\neq t}^{\NToda} \biggl[\frac{\Gamma_b(b+b^{-1}+\vev{\I a_{12},h_s-h_t})}{\Gamma_b(\vev{\I a_{32},h_s-h_t})}\biggr]
    \\
    & \times \prod_{s,t=1}^{\NToda} \biggl[
    \frac{\Gamma_b(\frac{\kappa_2}{\NToda}+\vev{\I a_3,h_s}-\vev{\I a_{32},h_t})
      \Gamma_b(b+b^{-1}-\frac{\kappa_4}{\NToda}-\vev{\I a_1,h_s}-\vev{\I a_{32},h_t})}
    {\Gamma_b(\frac{\kappa_2}{\NToda}+\vev{\I a_1,h_s}-\vev{\I a_{12},h_t})
      \Gamma_b(b+b^{-1}-\frac{\kappa_4}{\NToda}-\vev{\I a_3,h_s}-\vev{\I a_{12},h_t})}
    \biggr]
    \\
    & \times
    \int \frac{\dd[^{\NToda-1}]{\tau}}{\prod_{i\neq j}^{\NToda-1} S_b(\I\tau_i-\I\tau_j)}
    \prod_{j=1}^{\NToda-1} \prod_{s=1}^{\NToda} \biggl[
      \frac{S_b(-\vev{\I a_3,h_s}+\I\tau_j) S_b(\frac{\kappa_2+\kappa_4}{\NToda}-b-b^{-1}+\vev{\I a_1,h_s}+\I\tau_j)}
      {S_b(\frac{\kappa_2}{\NToda}-\vev{\I a_{32},h_s}+\I\tau_j) S_b(\frac{\kappa_4}{\NToda}+\vev{\I a_{12},h_s}+\I\tau_j)}
      \biggr]
  \end{aligned}
\end{equation}
up to a constant factor that does not depend on any momentum.  The integration contours go
from $-\infty$ to $\infty$ with poles of the numerator~$S_b$ functions above the contours,
and zeros of the denominator below them.  For instance, if all components $\vev{\I
a_1,h_s}$, $\vev{\I a_3,h_2}$, $\vev{\I a_{12},h_2}$, $\vev{\I a_{32},h_2}$ are purely
imaginary and all $\Re\frac{\kappa_i}{N}=(b+b^{-1})/2$ then contours can be taken to be
horizontal lines with $-(b+b^{-1})/2 < \Im(\tau_j) < 0$.
For other values of momenta, the contour is deformed to keep the same set of poles on each
side.  Another remark is that
$\prod_{i\neq j}^{\NToda-1} 1/S_b(\I\tau_i-\I\tau_j)$ has no pole: it simplifies
to a product of sines~\eqref{kernel-Sb-to-sines}.

In a normalization of conformal blocks where the leading term is a power of~$x$, the
braiding kernel includes phases, depending on the sign $\epsilon$ of $\Im x$:
\begin{equation}\label{kernel-phases}
  \Braiding{\epsilon}{\alpha_{12}\alpha_{32}}{\kappa_4 h_1\! & \!\kappa_2 h_1\\\alpha_3 & \alpha_1}
  = e^{\I\pi\epsilon[\dimToda(\alpha_{12})-\dimToda(\alpha_1)+\dimToda(\alpha_{32})-\dimToda(\alpha_3)]}
  \Braiding{}{\alpha_{12}\alpha_{32}}{\kappa_4 h_1\! & \!\kappa_2 h_1\\\alpha_3 & \alpha_1}
\end{equation}

A preliminary check of~\eqref{kernel-expression} is that it reproduces known
results~\cite{Ponsot:1999uf} for the Liouville theory ($\NToda=2$).  In their
equation~(48) replace their $Q$ by $b+b^{-1}$, shift the integration variable
$s\to s-\alpha_{21}+\alpha_4-(b+b^{-1})/2$, then map $\alpha_2\to b+b^{-1}-\alpha_2$ (for $\NToda=2$ this
is a Weyl symmetry).  The factors with $U_{3,4}$ become
$S_b\bigl(\pm(\alpha_3-(b+b^{-1})/2)+s\bigr)$.  The factors with $U_{1,2}$ become
$S_b\bigl(\alpha_4+\alpha_2-b-b^{-1}\pm(\alpha_1-(b+b^{-1})/2)+s\bigr)$.  The denominator factors with
$\widehat{V}_{1,2}$ become $S_b\bigl(\alpha_2\pm(\alpha_{32}-(b+b^{-1})/2)+s\bigr)$.  The
denominator factors with $\widehat{V}_{3,4}$ become
$S_b\bigl(\alpha_4\pm(\alpha_{21}-(b+b^{-1})/2)+s\bigr)$.  Thus, the integrand
from~\cite{Ponsot:1999uf,Ponsot:2000mt}
coincides with that of~\eqref{kernel-expression} for $\NToda=2$.
It is straightforward to check that prefactors also coincide.

\subsection{Reduction to fully degenerate}
\label{ssec:kernel-degenerate}

We now describe how to take the limit $\kappa_2 h_1\to -Kbh_1$
in~\eqref{kernel-expression}, and retrieve the sum of residues from
\autoref{ssec:matrices-symmetric}.

The integrand in~\eqref{kernel-expression} has poles at
\begin{equation}\label{kernel-degenerate-lpoles}
  \I\tau_j =
  \begin{cases}
    \vev{\I a_3,h_s} - mb - n/b \\
    b+b^{-1}-\frac{\kappa_2}{\NToda}-\frac{\kappa_4}{\NToda}-\vev{\I a_1,h_s} - mb - n/b
  \end{cases}
\end{equation}
and
\begin{equation}\label{kernel-degenerate-rpoles}
  \I\tau_j =
  \begin{cases}
    b+b^{-1}-\frac{\kappa_2}{\NToda}+\vev{\I a_{32},h_s} + mb + n/b \\
    b+b^{-1}-\frac{\kappa_4}{\NToda}-\vev{\I a_{12},h_s} + mb + n/b
  \end{cases}
\end{equation}
for integers $m,n\geq 0$.  As mentionned before, the contour for $\I\tau$ is chosen with
poles~\eqref{kernel-degenerate-lpoles} on the left and
poles~\eqref{kernel-degenerate-rpoles} on the right.  This is possible as long as the two
sets of poles are disjoint.  Otherwise, the contour is pinched between the two sets and
the integral diverges.

To understand the divergence, consider a simple model of a contour integral pinched by
poles getting close together from the two sides of the contour.  Let $f(z)$ be holomorphic
in a neighborhood of~$a$, and $a_L$ and $a_R$ be points in this neighborhood.  Then
\begin{equation}
  \int_{\text{between}} \dd{z} \frac{f(z)}{(z-a_L)(z-a_R)}
  = 2\pi\I \frac{f(a_L)}{a_L-a_R} + \int_{\text{left}} \dd{z} \frac{f(z)}{(z-a_L)(z-a_R)}
\end{equation}
where the initial contour goes between the two points, with $a_L$ on its left and $a_R$ on
its right, and where the second contour is moved through~$a_L$.  As $a_L,a_R\to a$, the
second term is regular, so the residue is $2\pi\I f(a)$.  This residue is obtained from
the original integrand by taking the limit $a_L,a_R\to a$ then considering the second
residue of the result $f(z)/(z-a)^2$.  We denote this operation of taking the second
residue as ${\res}^2$.

In our case there are two rather different pinchings by
poles~\eqref{kernel-degenerate-lpoles} and~\eqref{kernel-degenerate-rpoles}.  Whenever one
of the~\eqref{kernel-degenerate-lpoles} is equal to
$b+b^{-1}-\frac{\kappa_2}{\NToda}+\vev{\I a_{32},h_s}+mb+n/b$ the braiding kernel is
singular.  In fact, these singularities, together with those of numerator
$\Gamma_b$~functions in~\eqref{kernel-expression}, precisely reproduce singularities of
the u-channel Toda CFT three-point functions:
\begin{equation}
  \begin{aligned}
    & \widehat{C}(\alpha_3,\kappa_2 h_1,2Q-\alpha_{32}) \widehat{C}(\alpha_{32},\kappa_4 h_1,\alpha_1)
    \\
    & = \frac{1}{\prod_{t,u} \bigl[ \Upsilon_b(\frac{\kappa_2}{\NToda}+\vev{\I a_3,h_t}-\vev{\I a_{32},h_u})
      \Upsilon_b(\frac{\kappa_4}{\NToda}+\vev{\I a_1,h_t}+\vev{\I a_{32},h_u}) \bigr]} \,.
  \end{aligned}
\end{equation}
It may be interesting to pursue further the analysis by considering multiple
singularities, keeping in mind the constraints $\sum_t \vev{\I a_j,h_t} = 0$ for each
momentum.  On the other hand, if one of the~\eqref{kernel-degenerate-lpoles} is equal to
$b+b^{-1}-\frac{\kappa_4}{\NToda}-\vev{\I a_{12},h_s}+mb+n/b$, the contour is also
pinched, but the prefactors in~\eqref{kernel-expression} (specifically the denominator
$\Gamma_b$~functions) cancel the singularity so the braiding kernel can have a finite
limit.

We are ready to consider our limit of interest: $\kappa_2 = -Kb + \NToda\I\varepsilon$ for
$\varepsilon\to 0$ (and $\varepsilon>0$).  The OPE of $\widehat{V}_{-Kbh_1}$ with a
generic vertex operator constrains $\alpha_{12}$ and~$\alpha_{32}$, so we further focus on
$\alpha_{12} = \alpha_1 - bh_{[n]}$ and $\alpha_{32} = \alpha_3 - bh_{[\anti{n}]}$ for
some weights $h_{[n]}$ and~$h_{[\anti{n}]}$ in $\repr(Kh_1)$.  To simplify some
expressions we write $\kappa_4 = \varkappa + Kb$.  We keep $\alpha_1$ and $\alpha_3$
generic.  The poles~\eqref{kernel-degenerate-lpoles} and~\eqref{kernel-degenerate-rpoles}
are now respectively at
\begin{align}
  \I\tau_j & =
  \begin{cases}
    \vev{\I a_3,h_s} - mb - n/b , \\
    b+b^{-1}-\frac{\varkappa}{\NToda}-\vev{\I a_1,h_s} - mb - n/b -\I\varepsilon ,
  \end{cases}
  \\
  \I\tau_j & =
  \begin{cases}
    b+b^{-1}+\vev{\I a_3,h_s}+b\anti{n}_s + mb + n/b -\I\varepsilon , \\
    b+b^{-1}-\frac{\varkappa}{\NToda}-\vev{\I a_1,h_s} - bn_s + mb + n/b \,.
  \end{cases}
\end{align}
Because the momenta $\alpha_1$ and $\alpha_3$ are generic, the contour is pinched as
$\varepsilon\to 0$ precisely when
$\I\tau_j = b+b^{-1}-\frac{\varkappa}{\NToda}-\vev{\I a_1,h_s} - bl$ for any
$1\leq j\leq\NToda-1$, $1\leq s\leq\NToda$ and $0\leq l\leq n_s$.  The most singular
contribution to the integral, of order $1/\varepsilon^{\NToda-1}$, comes from values
of~$\I\tau$ where all~$\I\tau_j$ take this form.

We only describe the contour integral part of the braiding
matrix~\eqref{kernel-expression}, as prefactors only lengthen computations.  The
coefficient of the term of order $1/\varepsilon^{\NToda-1}$ in this integral is
\begin{equation}
  \begin{aligned}
    I & = \prod_{j=1}^{\NToda-1} \biggl[
    \sum_{p_j=1}^{\NToda} \sum_{k_j=0}^{n_{p_j}} \mathop{{\res}^2}\limits_{\I\tau_j = b+b^{-1}-\frac{\varkappa}{\NToda}-\vev{\I a_1,h_{p_j}} - k_j b}
    \biggr]
    \Biggl\{ \prod_{i\neq j}^{\NToda-1} \frac{1}{S_b(\I\tau_i-\I\tau_j)} \\
    & \quad \times \prod_{j=1}^{\NToda-1} \prod_{s=1}^{\NToda} \biggl[
      \frac{S_b(-\vev{\I a_3,h_s}+\I\tau_j)}{S_b(-\vev{\I a_3,h_s}-\anti{n}_sb+\I\tau_j)}
      \frac{S_b(\frac{\varkappa}{\NToda}-b-b^{-1}+\vev{\I a_1,h_s}+\I\tau_j)}{S_b(\frac{\varkappa}{\NToda}+\vev{\I a_1,h_s}+n_sb+\I\tau_j)}
      \biggr]
      \Biggr\} \,.
  \end{aligned}
\end{equation}
Note that
\begin{equation}\label{kernel-Sb-to-sines}
  \prod_{i\neq j}^{\NToda-1} \frac{1}{S_b(\I\tau_i-\I\tau_j)}
  = \prod_{i<j}^{\NToda-1} \Bigl(- 4 \sin\pi b(\I\tau_i-\I\tau_j) \sin\frac{\pi}{b}(\I\tau_i-\I\tau_j) \Bigr) \,.
\end{equation}
The shift relations for $\Gamma_b$ and~$S_b$ yield
\begin{equation}
  \begin{aligned}
    I & = \prod_{j=1}^{\NToda-1} \biggl[
    \sum_{p_j=1}^{\NToda} \sum_{k_j=0}^{n_{p_j}} \mathop{{\res}^2}\limits_{\I\tau_j = b+b^{-1}-\frac{\varkappa}{\NToda}-\vev{\I a_1,h_{p_j}} - k_j b}
    \biggr]
    \Biggl\{ \\
    & \qquad \frac{\prod_{i<j}^{\NToda-1} \bigl[- 4 b^2 \sin\pi b(\I\tau_i-\I\tau_j) \sin\frac{\pi}{b}(\I\tau_i-\I\tau_j) \bigr]}
    {\prod_{j=1}^{\NToda-1} \prod_{s=1}^{\NToda} \bigl[2\sin\frac{\pi}{b}(\frac{\varkappa}{\NToda}-b-b^{-1}+\vev{\I a_1,h_s}+\I\tau_j)\bigr]}
    \\
    & \qquad
    \times \prod_{j=1}^{\NToda-1} \prod_{s=1}^{\NToda} \biggl[
    \frac{\prod_{\mu=0}^{\anti{n}_s-1} \bigl[2\sin\pi(-b^2-\mu b^2-b\vev{\I a_3,h_s}+b\I\tau_j)\bigr]}
    {\prod_{\mu=0}^{n_s} \bigl[2\sin\pi(-b^2+\mu b^2+\frac{b\varkappa}{\NToda}+b\vev{\I a_1,h_s}+b\I\tau_j)\bigr]}
      \biggr]
      \Biggr\} \,.
  \end{aligned}
\end{equation}
This expression differs from the desired sum of residues~\eqref{matrices-sym-expr} in the
following respects: a redefinition
$\I\tau_j\to b\I\tau_j+\frac{b\varkappa}{\NToda}-b^2-1$, a sum over choices of the~$p_j$,
and additional factors of the form $\sin\frac{\pi}{b}(\cdots)$.  Because of
antiperiodicity of sine these factors are independent of the~$k_j$ except for a sign.
After extracting a sign, these factors are
\begin{equation}
  \frac{\prod_{i<j}^{\NToda-1} \sin\frac{\pi}{b}(\vev{\I a_1,h_{p_i}-h_{p_j}})}
    {\prod_{j=1}^{\NToda-1} \prod_{s\neq p_j}^{\NToda} \sin\frac{\pi}{b}(\vev{\I a_1,h_s-h_{p_j}})}
  =
  \begin{cases}
    0 \quad \text{if two $p_i$ coincide, otherwise} \\
    1\Big/ \prod_{j=1}^{\NToda-1} \prod_{i=j+1}^{\NToda} \sin\frac{\pi}{b}(\vev{\I a_1,h_{p_i}-h_{p_j}}) \,,
  \end{cases}
\end{equation}
where $p_{\NToda}$ denotes the (single) element of
$\intset{1}{\NToda}\setminus\{p_i\mid i<\NToda\}$ so that $p$ is a permutation of
$\intset{1}{\NToda}$.  Then these factors are independent of the permutation~$p$, except
for a sign: the signature of~$p$.  For each permutation~$p$ we get a sum of residues times
the signature of~$p$, and this structure coincides with that of~\eqref{matrices-sym-expr}.
Below that equation we had proven that it is independent of the permutation, hence summing
over permutation simply introduces a trivial factor.  We have thus reproduced
qualitatively the structure of the braiding matrix of~$\widehat{V}_{-Kbh_1}$ by taking the
appropriate limit of the braiding kernel.  The exact reduction of the braiding kernel to
the braiding matrix is confirmed by a more detailed calculation.

\subsection{Shift relation from pentagon identity}
\label{ssec:kernel-shift}

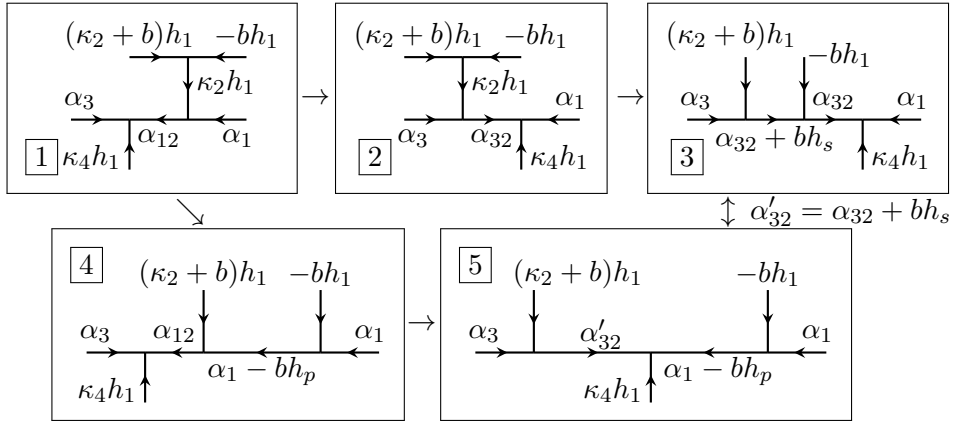
\begin{figure}
  \centering
  \begin{tikzpicture}
    \node at (0,0) {\fbox{%
      \mathtikz[x=1em,y=1ex,thick]{
        \draw[->-=.55] (0,0) -- (2,0);
        \draw[->-=.55] (6,0) -- (4,0);
        \draw[->-=.55] (4,0) -- (2,0);
        \draw[->-=.55] (2,-4) -- (2,0);
        \draw[->-=.55] (4,5) -- (4,0);
        \draw[->-=.55] (2,5) -- (4,5);
        \draw[->-=.55] (6,5) -- (4,5);
        \node at (0.3,1.5) {$\alpha_3$};
        \node at (5.7,-1.5) {$\alpha_1$};
        \node at (3,-1.5) {$\alpha_{12}$};
        \node at (5.3,3) {$\kappa_2 h_1$};
        \node at (6,6.7) {$-bh_1$};
        \node at (2,6.7) {$(\kappa_2+b)h_1$};
        \node at (0.7,-3) {$\kappa_4 h_1$};
        \node at (-1,-3) {\fbox{$1$}};
      }%
    }};
    \node at (2.15,0) {$\rightarrow$};
    \node at (4.2,0) {\fbox{%
      \mathtikz[x=1em,y=1ex,thick]{
        \draw[->-=.55] (0,0) -- (2,0);
        \draw[->-=.55] (6,0) -- (4,0);
        \draw[->-=.55] (2,0) -- (4,0);
        \draw[->-=.55] (4,-4) -- (4,0);
        \draw[->-=.55] (2,5) -- (2,0);
        \draw[->-=.55] (0,5) -- (2,5);
        \draw[->-=.55] (4,5) -- (2,5);
        \node at (0.3,-1.5) {$\alpha_3$};
        \node at (5.7,1.5) {$\alpha_1$};
        \node at (3,-1.5) {$\alpha_{32}$};
        \node at (3.3,3) {$\kappa_2 h_1$};
        \node at (0.5,6.7) {$(\kappa_2+b)h_1$};
        \node at (4.5,6.7) {$-bh_1$};
        \node at (5.3,-3) {$\kappa_4 h_1$};
        \node at (-1,-3) {\fbox{$2$}};
      }%
    }};
    \node at (6.25,0) {$\rightarrow$};
    \node at (8.5,0) {\fbox{%
      \mathtikz[x=1em,y=1ex,thick]{
        \draw[->-=.55] (0,0) -- (2,0);
        \draw[->-=.55] (2,0) -- (4,0);
        \draw[->-=.55] (4,0) -- (6,0);
        \draw[->-=.55] (8,0) -- (6,0);
        \draw[->-=.55] (6,-4) -- (6,0);
        \draw[->-=.55] (2,5) -- (2,0);
        \draw[->-=.55] (4,5) -- (4,0);
        \node at (0.3,1.5) {$\alpha_3$};
        \node at (7.7,1.5) {$\alpha_1$};
        \node at (5,1.5) {$\alpha_{32}$};
        \node at (3,-1.5) {$\alpha_{32}+bh_s$};
        \node at (5.2,5.2) {$-bh_1$};
        \node at (1.5,6.7) {$(\kappa_2+b)h_1$};
        \node at (7.3,-3) {$\kappa_4 h_1$};
        \node at (0,-3) {\fbox{$3$}};
      }%
    }};
    \node at (0.5,-1.55) {$\searrow$};
    \node at (1,-3.1) {\fbox{%
      \mathtikz[x=1em,y=1ex,thick]{
        \draw[->-=.55] (0,0) -- (2,0);
        \draw[->-=.55] (10,0) -- (8,0);
        \draw[->-=.55] (8,0) -- (4,0);
        \draw[->-=.55] (4,0) -- (2,0);
        \draw[->-=.55] (2,-4) -- (2,0);
        \draw[->-=.55] (8,5) -- (8,0);
        \draw[->-=.55] (4,5) -- (4,0);
        \node at (0.3,1.5) {$\alpha_3$};
        \node at (9.7,1.5) {$\alpha_1$};
        \node at (6,-1.5) {$\alpha_1-bh_p$};
        \node at (3,1.5) {$\alpha_{12}$};
        \node at (8,6.3) {$-bh_1$};
        \node at (4,6.3) {$(\kappa_2+b)h_1$};
        \node at (0.7,-3) {$\kappa_4 h_1$};
        \node at (0,7) {\fbox{$4$}};
      }%
    }};
    \node at (3.55,-3.1) {$\rightarrow$};
    \node at (6.5,-3.1) {\fbox{%
      \mathtikz[x=1em,y=1ex,thick]{
        \draw[->-=.55] (0,0) -- (2,0);
        \draw[->-=.55] (2,0) -- (6,0);
        \draw[->-=.55] (12,0) -- (10,0);
        \draw[->-=.55] (10,0) -- (6,0);
        \draw[->-=.55] (6,-4) -- (6,0);
        \draw[->-=.55] (10,5) -- (10,0);
        \draw[->-=.55] (2,5) -- (2,0);
        \node at (0.3,1.5) {$\alpha_3$};
        \node at (11.7,1.5) {$\alpha_1$};
        \node at (8.3,-1.5) {$\alpha_1-bh_p$};
        \node at (4.3,1.5) {$\alpha'_{32}$};
        \node at (10,6.3) {$-bh_1$};
        \node at (3.5,6.3) {$(\kappa_2+b)h_1$};
        \node at (4.7,-3) {$\kappa_4 h_1$};
        \node at (0,7) {\fbox{$5$}};
      }%
    }};
    \node at (9,-1.55) {$\updownarrow\:\alpha'_{32}=\alpha_{32}+bh_s$};
  \end{tikzpicture}
  \caption{\label{fig:kernel-shift-pentagon}%
    Pentagon identity. $1\to 2$ and $4\to 5$ are braidings of two semi-degenerates,
    $2\to 3$ and $1\to 4$ are known fusions of $\widehat{V}_{-bh_1}$ and a
    semi-degenerate, $5\leftrightarrow 3$ is a known braiding of $\widehat{V}_{-bh_1}$
    and a semi-degenerate.}
\end{figure}

Braiding and fusion kernels (or matrices) obey Moore--Seiberg pentagon and hexagon
relations.  Here we consider a particular pentagon relation shown in
\autoref{fig:kernel-shift-pentagon}.  Going through the moves $1\to 2\to 3$ we find
\begin{align}
  \Fblock{}{}{\nomatrix 1}{}
  & = \int \dd{\alpha_{32}}
  \Braiding{}{\alpha_{12}\alpha_{32}}{\kappa_4 h_1 & \kappa_2 h_1 \\ \alpha_3 & \alpha_1}
  \Fblock{}{}{\nomatrix 2}{}
  \\\label{kernel-shift-123}
  & = \int \dd{\alpha_{32}} \sum_{s=1}^{\NToda}
  \Braiding{}{\alpha_{12}\alpha_{32}}{\kappa_4 h_1 & \kappa_2 h_1 \\ \alpha_3 & \alpha_1}
  \Fusion{}{s}{(\kappa_2+b)h_1 & -bh_1 \\ \alpha_3 & 2Q-\alpha_{32}}
  \Fblock{}{}{\nomatrix 3}{} \,.
\end{align}
On the other hand, going through the moves $1\to 4\to 5\to 3$ yields
\begin{align}
  & \Fblock{}{}{\nomatrix 1}{}
  = \sum_{p=1}^{\NToda}
  \Fusion{}{p}{(\kappa_2+b)h_1 & -bh_1 \\ 2Q-\alpha_{12} & \alpha_1}
  \Fblock{}{}{\nomatrix 4}{}
  \displaybreak[0]\\
  & = \sum_{p=1}^{\NToda} \int \dd{\alpha'_{32}}
  \Fusion{}{p}{(\kappa_2+b)h_1 & -bh_1 \\ 2Q-\alpha_{12} & \alpha_1}
  \Braiding{}{\alpha_{12}\alpha'_{32}}{\kappa_4 h_1 & (\kappa_2+b)h_1 \\ \alpha_3 & \alpha_1-bh_p}
  \Fblock{}{}{\nomatrix 5}{}
  \displaybreak[0]\\\label{kernel-shift-1453}
  &
    \begin{aligned}
      = \sum_{p,s=1}^{\NToda} \int \dd{\alpha'_{32}} \,
      & \Fusion{}{p}{(\kappa_2+b)h_1 & -bh_1 \\ 2Q-\alpha_{12} & \alpha_1}
      \Braiding{}{\alpha_{12}\alpha'_{32}}{\kappa_4 h_1 & (\kappa_2+b)h_1 \\ \alpha_3 & \alpha_1-bh_p}
       \Braiding{}{ps}{\kappa_4 h_1 & -bh_1 \\ \alpha'_{32} & \alpha_1}
      \Fblock{}{}{\nomatrix 3}{} \,.
    \end{aligned}
\end{align}
The coefficients of each conformal block $\Fblock{}{}{\nomatrix 3}{}$ (these are labelled
by the choice of $1\leq s\leq\NToda$ and $\alpha_{32}=\alpha'_{32}-bh_s$) must be the same
in \eqref{kernel-shift-123} and~\eqref{kernel-shift-1453}.

To check that the proposal~\eqref{kernel-expression} obeys the pentagon identity, we will
need the braiding matrix obtained from~\eqref{matrices-fund-expr} using
$\alpha_1 = Q - \I a_1$ and $\alpha'_{32} = Q-\I a_{32}+bh_s$:
\begin{equation}
  \begin{aligned}
    & \Braiding{}{ps}{\kappa_4 h_1 & -bh_1 \\ \alpha'_{32} & \alpha_1} \\
    & \quad = 
    \prod_{t\neq p}^{\NToda} \frac{\Gamma(1 + b\vev{\I a_1,h_t-h_p})}{\Gamma(\frac{b\kappa_4}{\NToda}+b\vev{\I a_1,h_t}+b\vev{\I a_{32},h_s}-b^2)}
    \\
    & \qquad \times
    \prod_{u\neq s}^{\NToda} \frac{\Gamma(b\vev{\I a_{32},h_s-h_u}-b^2)}{\Gamma(1 - \frac{b\kappa_4}{\NToda}-b\vev{\I a_1,h_p}-b\vev{\I a_{32},h_u})}
    \,.
  \end{aligned}
\end{equation}
We will also need coefficients of the fusion of $(\kappa_2+b)h_1$ and $-bh_1$ into
$\kappa_2 h_1$, which can be deduced from the braiding matrix~\eqref{matrices-fund-expr},
as done in equation~(B.14) of the reference \cite{Gomis:2010kv}.
\begin{equation}
  \Fusion{}{p}{(\kappa_2+b)h_1 & -bh_1 \\ Q-\I a' & Q-\I a}
  = \Gamma(b\kappa_2)
  \frac{\prod_{t\neq p}^{\NToda}\Gamma(b\vev{\I a,h_p-h_t})}
  {\prod_{t=1}^{\NToda}\Gamma(\frac{b\kappa_2}{\NToda}+b\vev{\I a,h_p}+b\vev{\I a',h_t})} \,.
\end{equation}

We now write down~\eqref{kernel-shift-1453} explicitly for a fixed choice of~$\alpha_{32}$
and of $1\leq s\leq\NToda$, and simplify it in order to find~\eqref{kernel-shift-123}.
All generic momenta are written as $\alpha=Q-\I a$ and we denote $\I a^u = \vev{\I a,h_u}$
for conciseness.  Note that $\alpha'_{32}=Q-\I a_{32}+bh_s$.

Let us start!
\begin{equation}
  \begin{aligned}
    & \sum_{p=1}^{\NToda}
    \Fusion{}{p}{(\kappa_2+b)h_1 & -bh_1 \\ 2Q-\alpha_{12} & \alpha_1}
    \Braiding{}{\alpha_{12}\alpha'_{32}}{\kappa_4 h_1 & (\kappa_2+b)h_1 \\ \alpha_3 & \alpha_1-bh_p}
    \Braiding{}{ps}{\kappa_4 h_1 & -bh_1 \\ \alpha'_{32} & \alpha_1}
    \\
    & = \sum_{p=1}^{\NToda} \Biggl(
    \Gamma(b\kappa_2)
    \frac{\prod_{t\neq p}^{\NToda}\Gamma(b\I a_1^p-b\I a_1^t)}
    {\prod_{t=1}^{\NToda}\Gamma(\frac{b\kappa_2}{\NToda}+b\I a_1^p-b\I a_{12}^t)}
    \prod_{t\neq u}^{\NToda} \frac{\Gamma_b(b+b^{-1}+\I a_{12}^t-\I a_{12}^u)}{\Gamma_b(\I a_{32}^t-\I a_{32}^u-b\delta_{s,t}+b\delta_{s,u})}
    \\
    & \quad \times
    \prod_{t,u=1}^{\NToda} \biggl[
    \frac{\Gamma_b(\frac{\kappa_2}{\NToda}+\I a_3^t-\I a_{32}^u+b\delta_{s,u})
      \Gamma_b(b+b^{-1}-\frac{\kappa_4}{\NToda}-\I a_1^t-\I a_{32}^u-b\delta_{p,t}+b\delta_{s,u})}
    {\Gamma_b(\frac{\kappa_2}{\NToda}+\I a_1^t-\I a_{12}^u+b\delta_{p,t})
      \Gamma_b(b+b^{-1}-\frac{\kappa_4}{\NToda}-\I a_3^t-\I a_{12}^u)}
    \biggr] \\
    & \quad \times
    \int \frac{\dd[^{\NToda-1}]{\tau_j}}{\prod_{i\neq j}^{\NToda-1} S_b(\I\tau_i-\I\tau_j)}
    \prod_{j=1}^{\NToda-1} \prod_{t=1}^{\NToda}
    \frac{S_b(-\I a_3^t+\I\tau_j) S_b(\frac{\kappa_2}{\NToda}+\frac{\kappa_4}{\NToda}-b-b^{-1}+\I a_1^t+\I\tau_j+b\delta_{p,t})}
    {S_b(\frac{\kappa_2}{\NToda}-\I a_{32}^t+\I\tau_j+b\delta_{s,t}) S_b(\frac{\kappa_4}{\NToda}+\I a_{12}^t+\I\tau_j)}
    \\
    & \quad \times
    \prod_{t\neq p}^{\NToda}
    \frac{\Gamma(1 + b\I a_1^t-b\I a_1^p)}{\Gamma(\frac{b\kappa_4}{\NToda}+b\I a_1^t+b\I a_{32}^s-b^2)}
    \prod_{u\neq s}^{\NToda}
    \frac{\Gamma(b\I a_{32}^s-b\I a_{32}^u-b^2)}{\Gamma(1 - \frac{b\kappa_4}{\NToda}-b\I a_1^p-b\I a_{32}^u)}
    \Biggr) \,.
  \end{aligned}
\end{equation}
We collect factors which do not depend on $p,s$ using shift relations of $\Gamma_b$
and~$S_b$.  Factors of $\sqrt{2\pi}$ and powers of~$b$ cancel, and we combine many Gamma as
$\Gamma(x)\Gamma(1-x)=\pi/\sin\pi x$.
\begin{equation}\label{kernel-shift-1453-c}
  \begin{aligned}
    & = \prod_{t,u=1}^{\NToda} \biggl[
    \frac{\Gamma_b(\frac{\kappa_2}{\NToda}+\I a_3^t-\I a_{32}^u)
      \Gamma_b(b+b^{-1}-\frac{\kappa_4}{\NToda}-\I a_1^t-\I a_{32}^u)}
    {\Gamma_b(\frac{\kappa_2}{\NToda}+\I a_1^t-\I a_{12}^u)
      \Gamma_b(b+b^{-1}-\frac{\kappa_4}{\NToda}-\I a_3^t-\I a_{12}^u)}
    \biggr]
    \prod_{t\neq u}^{\NToda} \biggl[\frac{\Gamma_b(b+b^{-1}+\I a_{12}^t-\I a_{12}^u)}{\Gamma_b(\I a_{32}^t-\I a_{32}^u)}\biggr]
    \\
    & \quad \times
    \int \frac{\dd[^{\NToda-1}]{\tau_j}}{\prod_{i\neq j}^{\NToda-1} S_b(\I\tau_i-\I\tau_j)} \Biggl\{
    \prod_{j=1}^{\NToda-1} \prod_{t=1}^{\NToda} \biggl[
    \frac{S_b(-\I a_3^t+\I\tau_j) S_b(\frac{\kappa_2}{\NToda}+\frac{\kappa_4}{\NToda}-b-b^{-1}+\I a_1^t+\I\tau_j)}
    {S_b(\frac{\kappa_2}{\NToda}-\I a_{32}^t+\I\tau_j) S_b(\frac{\kappa_4}{\NToda}+\I a_{12}^t+\I\tau_j)}
    \biggr]
    \\
    & \quad \qquad \times
    \frac{
      \Gamma(b\kappa_2)\prod_{u\neq s}^{\NToda} \Gamma(b\I a_{32}^u-b\I a_{32}^s)
      \prod_{t=1}^{\NToda} \frac{1}{\pi}\sin\pi(\frac{b\kappa_4}{\NToda}+b\I a_1^t+b\I a_{32}^s-b^2)
    }
    {
      \prod_{j=1}^{\NToda-1} \frac{1}{\pi}\sin\pi b(\frac{\kappa_2}{\NToda}-\I a_{32}^s+\I\tau_j)
      \prod_{t=1}^{\NToda} \Gamma(\frac{b\kappa_2}{\NToda}+b\I a_3^t-b\I a_{32}^s)
    }
    \\
    & \quad \qquad \times
    \sum_{p=1}^{\NToda}
    \frac{
      \prod_{j=1}^{\NToda-1} \frac{1}{\pi}\sin\pi b(\frac{\kappa_2}{\NToda}+\frac{\kappa_4}{\NToda}-b-b^{-1}+\I a_1^p+\I\tau_j)
    }
    {
      \frac{1}{\pi}\sin\pi(\frac{b\kappa_4}{\NToda}+b\I a_1^p+b\I a_{32}^s-b^2)
      \prod_{t\neq p}^{\NToda} \frac{1}{\pi}\sin\pi(b\I a_1^p-b\I a_1^t)
    }
    \Biggr\}
  \end{aligned}
\end{equation}
The last line is a sum of residues at $\upsilon=b\I a_1^p$ of
$\prod_{j=1}^{\NToda-1} \bigl[\frac{1}{\pi}\sin\pi
(\frac{b\kappa_2}{\NToda}+\frac{b\kappa_4}{\NToda}-b^2-1+\upsilon+b\I\tau_j) \bigr] \bigm/
\bigl[\frac{1}{\pi}\sin\pi(\frac{b\kappa_4}{\NToda}+\upsilon+b\I a_{32}^s-b^2)
\prod_{t=1}^{\NToda} \frac{1}{\pi}\sin\pi(\upsilon-b\I a_1^t) \bigr]$,
which is equal to minus its residue at the last pole
$\upsilon=b^2-\frac{b\kappa_4}{\NToda}-b\I a_{32}^s$.  That residue turns out to cancel
most factors in the second to last line.  Together, these last two lines
of~\eqref{kernel-shift-1453-c} are equal to
\begin{equation}
  \frac{\Gamma(b\kappa_2)\prod_{u\neq s}^{\NToda} \Gamma(b\I a_{32}^u-b\I a_{32}^s)}
  {\prod_{t=1}^{\NToda} \Gamma(\frac{b\kappa_2}{\NToda}+b\I a_3^t-b\I a_{32}^s)}
  = \Fusion{}{s}{(\kappa_2+b)h_1 & -bh_1 \\ \alpha_3 & 2Q-\alpha_{32}} \,.
\end{equation}
In particular, this does not depend on~$\I\tau_j$ and can be pulled out of the integral.
The first two lines of~\eqref{kernel-shift-1453-c} then reproduce precisely the braiding
matrix~\eqref{kernel-expression} of two semi-degenerate vertex operators.  This concludes
our check of the pentagon relation $\eqref{kernel-shift-123}=\eqref{kernel-shift-1453}$.

This pentagon relation expresses the braiding
kernel~$\Braiding{}{\alpha_{12}\alpha_{32}}{}$ as a sum of~$\NToda$ braiding kernels with
$\kappa_2 \to \kappa_2+b$, $\alpha_1\to\alpha_1-bh_p$ and
$\alpha_{32}\to\alpha'_{32}=\alpha_{32}+bh_s$.  Thus, if the braiding kernel was known for
some value $\kappa_2=\lambda$, it could be deduced for $\kappa_2=\lambda-Kb$ for integer
$K\geq 0$.  The pentagon identity $(1\to 2\to 3\to 5) = (1\to 4\to 5)$ is checked through
very similar computations.  It allows the opposite shifts: from the $\kappa_2=\lambda$
braiding kernel one gets the $\kappa_2=\lambda+Kb$ braiding kernel.  By symmetry,
identical shift relations exist with $b\to\frac{1}{b}$, thus fixing braiding kernels for
$\kappa_2 = \lambda+Kb+L/b$ for all integers $K,L$.  For generic real~$b^2$, continuity
would then determine the braiding kernel uniquely.
Of course, this logic requires knowing for some $\kappa_2$ that the braiding kernel is
correct for all values of other momenta.

From the trivial braiding kernel at $\kappa_2=0$ the shift relations allow us to reach all
degenerate momenta $\kappa_2=-Kb-Lb^{-1}$ for $K,L\geq 0$, thus proving in particular that
the braiding matrix~\eqref{matrices-sym-expr} ($L=0$) is correct.  On the other hand,
operators $\widehat{V}_{-Kb-Lb^{-1}}$ vanish if $K=-1$ or $L=-1$, which prevents us from
using the pentagon relation to deduce prove our answer for $K<-1$ or $L<-1$.

To establish the proposed braiding kernel a strategy could be to derive shift relations on
each momentum separately as was done in the Liouville case.  We hope to return to this in
the future.  In the mean-time, the shift relations that we have already proven are at
least very strong evidence that the proposed braiding kernel is correct.

\section{Domain wall and its symmetries}
\label{sec:wall}

We want to describe the S-duality domain wall of 4d $\Nsusy=2$ $SU(N)$ SQCD as a 3d
$\Nsusy=2$ gauge theory coupled to the bulk on both sides.  As explained in
\autoref{sec:AGT}, its $S^3_b$ (ellipsoid) partition function is equal, up to
prefactors~\eqref{AGT-ZS3-CBZZ}, to the $W_N$~braiding kernel~\eqref{kernel-expression}
that we just computed.  To warm up we explain in \autoref{ssec:wall-theory} how continuous
flavour symmetries of the 4d/3d coupled systems given in the introduction reproduce those
expected of the S-duality wall.  Then we move on in \autoref{ssec:wall-Z} to more involved
expressions: comparing the braiding kernel to localization results.  In fact, these are
the formulas from which we read off the announced 3d $U(N-1)$ and $\USp(2N-2)$ gauge
theories.  \autoref{ssec:wall-duality} matches $\mathbb{Z}_2$~symmetries of the
$W_N$~braiding, of the abstract S-duality wall, and of its concrete 3d gauge theory
description.

Most interestingly, charge conjugation of 4d theories is not a manifest
symmetry of the 3d theory.  To show that our explicit expressions are invariant we lift
the 3d partition function of the $\USp(2N-2)$ theory to a 4d index then use Seiberg
duality for $\USp$ groups.  Alternatively, a physical understanding was obtained in~\cite{Benini:2017dud}
(after version 1 of this paper) where a variant of Aharony duality was found, under which
our $U(N-1)$ theory is self-dual.  Finally, we discuss peculiarities for $N=2$ due to
$\USp(2)=SU(2)$.

\subsection{Continuous flavour symmetries}
\label{ssec:wall-theory}

Before turning to partition functions and how we obtained the following descriptions we
explain how continuous flavour symmetries of the 4d/3d coupled systems reproduce those
expected of the S-duality wall.  The two quiver descriptions are (with notations below)
\begin{equation}\label{wall-ZZ}
  \vev{\text{S-duality wall}}
  =
  Z \!\left[
    \quiver{
      \node (1)   [color-group] {$\scriptstyle U(N-1)$};
      \node (R)   [flavor-group, right=1em of 1] {$2N$};
      \node (A)   [color-group, above=3ex of 1] {$N$};
      \node (B)   [color-group, below=3ex of 1] {$N$};
      \draw[->-=.55] (R) -- (1);
      \draw[->-=.7] (1) -- (A);
      \draw[->-=.7] (1) -- (B);
      \draw (A) -- (R) node [midway, above right=-4pt] {4d};
      \draw (B) -- (R) node [midway, below right=-4pt] {4d};
    }
  \right]
  =
  \lim_{\mu\to\pm\infty}
  Z \!\left[
    \quiver{
      \node (1)   [color-group] {$\scriptstyle \USp(2N-2)$};
      \node (R)   [flavor-group, right=2.5em of 1] {$2N$};
      \node (A)   [color-group, above=3ex of 1] {$N$};
      \node (B)   [color-group, below=3ex of 1] {$N$};
      \draw[->-=.55] (R) -- (1) node [pos=.7,below=-.8ex] {$-\mu$};
      \draw[->-=.7] (1) -- (A) node [midway,left] {$+\mu$};
      \draw[->-=.7] (1) -- (B) node [midway,left] {$+\mu$};
      \draw (A) -- (R) node [midway, above right=-4pt] {4d};
      \draw (B) -- (R) node [midway, below right=-4pt] {4d};
    }
  \right]
\end{equation}
with superpotential
\begin{equation}\label{wall-monopole-supo}
  W = \sum_{f=1}^{2N} \sum_{s=1}^{N} \Bigl( \Phi_{fs}\bigr|_{\text{3d}} \anti{q}_{s} q_{f} + \Phi'_{fs}\bigr|_{\text{3d}} \anti{q}'_{s} q_{f} \Bigr) +
  \begin{cases}
    V_+ + V_- & \text{for the $U(N-1)$ theory,} \\
    Y & \text{for the $\USp(2N-2)$ theory.}
  \end{cases}
\end{equation}
The quivers denote 4d/3d coupled systems involving a 3d $\Nsusy=2$ gauge theory with
$U(N-1)$ or $\USp(2N-2)$ gauge group, $2N$ fundamental chiral multiplets~$q_f$, $N+N$
antifundamentals $\anti{q}_s$, $\anti{q}'_s$, and a monopole superpotential made of all
minimal monopoles $V_{\pm}$ or~$Y$ for the given gauge group.  The $SU(N)$ flavour
symmetries of $\anti{q}$ and $\anti{q}'$ in this theory are each gauged using the 3d
restriction of the 4d $\Nsusy=2$ $SU(N)$ vector multiplet on one side of the wall.  The 3d
superpotential~\eqref{wall-monopole-supo} is built using the 3d $\Nsusy=2$ chiral
multiplets and using restrictions $\Phi|_{\text{3d}}$ and $\Phi'|_{\text{3d}}$ of the 4d
$\Nsusy=2$ hypermultiplets of both SQCD theories.  It is compatible with the $SU(N)^2$
gauging and breaks many other symmetries as explained next, identifying for example the
$SU(2N)$ flavour symmetries of~$q$ and of bulk hypermultiplets.

The $U(N-1)$ theory with $2N$ flavours has an
$SU(2N)_1\times SU(2N)_2\times U(1)_B\times U(1)_T$ flavour symmetry and $U(1)_R$
symmetry.  Gauging an $SU(N)^2$ subgroup of one $SU(2N)_1$ using the two 4d gauge
symmetries reduces that factor of the flavour symmetry to $U(1)_1$.  Besides being
consistent with this identification, the superpotential breaks many 3d and 4d symmetries
to their diagonal subgroup, thus identifying pairs of symmetries.
\begin{itemize}
\item The 3d symmetry $SU(2N)_2$ is identified to the $SU(2N)$ flavour symmetries of the
  two 4d theories by the first two terms in~$W$.  This reproduces the fact that S-duality
  identifies the $SU(2N)$ flavour symmetries of dual theories.
\item The first two terms in~$W$ further identify the $U(1)_1$ symmetry to the difference
  of baryonic $U(1)$ flavour symmetries of the two 4d theories, and the 3d baryonic
  symmetry $U(1)_B$ to their sum.
\item The dressed monopole operators $V_+$ and~$V_-$ have the same non-zero charge under
  the $U(1)_B$ flavour symmetry of the 3d theory and opposite charges under the
  topological one, so both $U(1)_B$ and $U(1)_T$ are broken by the superpotential terms
  $V_++V_-$.  In particular, only the difference of 4d baryonic symmetries survives
  (further combined with $U(1)_1$ as seen above).  This reproduces the fact that the
  baryonic symmetry is flipped by S-duality.
\end{itemize}
Furthermore, $W$~has $R$-charge~$2$ under the $U(1)_R$ symmetry of 3d $\Nsusy=2$.  This is
a subgroup of the $SU(2)_R$ symmetry of 4d $\Nsusy=2$, so $U(1)_R$-charges of the 4d
fields $\Phi$ and~$\Phi'$ are integers.  By continuity these charges on~$S^4_b$ must be
equal to those in flat space, which are~$1$ since hypermultiplet scalars are in a doublet
of $SU(2)_R$.  As a result, the $R$-charges of $q$ and~$\anti{q}$ must sum to~$1$, as do
those of $q$ and~$\anti{q}'$.  Up to a diagonal gauge redundancy, the superpotential thus
sets $R$-charges of all 3d chiral multiplets to their canonical UV value~$\frac{1}{2}$.
Since none of the remaining unbroken $U(1)$ symmetries leave all chiral multiplets
invariant, we learn that the UV and IR $R$-symmetries coincide.  A consistency check is
that under this UV $R$-symmetry of the 3d theory the monopole operators $V_{\pm}$ also
have the correct $R$-charge\footnote{Alternatively, one can start by analyzing the
  $U(N_c)$ theory with $N_f$ flavours and monopole superpotential, as done
  in~\cite{Benini:2017dud}.  The superpotential sets the $R$-charge of monopoles to~$2$,
  which fixes most possible mixing of $R$-symmetry with other symmetries, thus setting
  $R$-charges of $q$, $\anti{q}$, $\anti{q}'$ to their UV value.  In this approach, the
  consistency check is that the cubic superpotential coupling with $\Phi|_{\text{3d}}$ and
  $\Phi'|_{\text{3d}}$ has the correct $R$-charge~$2$.}.  Symmetries and charges are
summarized in \autoref{tab:symm}.

\begin{table}\centering
  \caption{\label{tab:symm}Charges of 3d and 4d fields under unbroken symmetries for
    the two descriptions of the S-duality domain wall in~\eqref{wall-ZZ}.  First we
    list gauge groups of the 3d~theory and 4d~theories, then flavour and $R$-symmetry
    groups.  Charges coincide except for the additional symmetry $U(1)_A$ in the second
    case, and the possibility to mix it into the $R$-symmetry for some parameter~$\nu$
    with $\abs{\nu}\leq 1/4$.}

  \begin{tabular}{c@{\,}c@{\,}c@{\,}c@{\,}c@{\,}c>{\quad}cc@{\,}c@{\,}c@{\,}c@{\,}c}
    \toprule
    & \multicolumn{5}{c}{Gauge symmetries} & & \multicolumn{5}{c}{Global symmetries} \\
    \cmidrule(lr){2-6}\cmidrule(lr){8-12}
    & $U(N-1)$ & $\times$ & $SU(N)_1$ & $\times$ & $SU(N)_2$ & & $U(1)_1$ & $\times$ & $SU(2N)_2$ & $\times$ & $U(1)_R$ \\
    \midrule
    $q$          & $\overline{N-1}$ & & $1$ & & $1$ & & $0$ & & $2N$ & & $\frac{1}{2}$ \\
    $\anti{q}$  & $N-1$ & & $\overline{N}$ & & $1$ & & $+1$ & & $1$ & & $\frac{1}{2}$ \\
    $\anti{q}'$ & $N-1$ & & $1$ & & $\overline{N}$ & & $-1$ & & $1$ & & $\frac{1}{2}$ \\
    $V_{\pm}$     & $1$ & & $1$ & & $1$ & & $0$ & & $1$ & & $2$ \\
    $\Phi$       & $1$ & & $N$ & & $1$ & & $-1$ & & $\overline{2N}$ & & $1$ \\
    $\Phi'$      & $1$ & & $1$ & & $N$ & & $+1$ & & $\overline{2N}$ & & $1$ \\
  \end{tabular}

  \begin{tabular}{c@{\,}c@{\,}c@{\,}c@{\,}c@{\,}c>{\quad}cc@{\,}c@{\,}c@{\,}c@{\,}c@{\,}c@{\,}c}
    \toprule
    & \multicolumn{5}{c}{Gauge symmetries} & & \multicolumn{7}{c}{Global symmetries} \\
    \cmidrule(lr){2-6}\cmidrule(lr){8-14}
    & $\USp(2N-2)$ & $\times$ & $SU(N)_1$ & $\times$ & $SU(N)_2$ & & $U(1)_1$ & $\times$ & $SU(2N)_2$ & $\times$ & $U(1)_A$ & $\times$ & $U(1)_R$ \\
    \midrule
    $q$          & $2N-2$ & & $1$ & & $1$ & & $0$ & & $2N$ & & $-1$ & & $\frac{1}{2}-\nu$ \\
    $\anti{q}$  & $2N-2$ & & $\overline{N}$ & & $1$ & & $+1$ & & $1$ & & $+1$ & & $\frac{1}{2}+\nu$ \\
    $\anti{q}'$ & $2N-2$ & & $1$ & & $\overline{N}$ & & $-1$ & & $1$ & & $+1$ & & $\frac{1}{2}+\nu$ \\
    $Y$          & $1$ & & $1$ & & $1$ & & $0$ & & $1$ & & $0$ & & $2$ \\
    $\Phi$       & $1$ & & $N$ & & $1$ & & $-1$ & & $\overline{2N}$ & & $0$ & & $1$ \\
    $\Phi'$      & $1$ & & $1$ & & $N$ & & $+1$ & & $\overline{2N}$ & & $0$ & & $1$ \\
    \bottomrule
  \end{tabular}
\end{table}

The $\USp(2N-2)$ theory is similar.  The fundamental and antifundamental representations
of $\USp$ are isomorphic, but gauging $SU(N)^2$ breaks the large flavour symmetry down to
$U(1)_1\times SU(2N)_2\times U(1)_B\times U(1)_A$,
where $U(1)_B$ and $U(1)_A$ act on $(q,\anti{q},\anti{q}')$ with charges
$(\pm 1,+1,+1)$.  The axial flavour symmetry $U(1)_A$ is new here because it was gauged in
the previous model.  As above, the superpotential identifies $SU(2N)_2$ with both $SU(2N)$
4d flavour symmetries, and identifies $U(1)_B$ and $U(1)_1$ with the sum and difference of
4d baryonic symmetries, respectively.  The monopole term~$Y$ is charged under $U(1)_B$
hence breaks this symmetry.  Fixing the $R$-charge of~$W$ to be~$2$ makes $q$,
$\anti{q}$, $\anti{q}'$ have canonical $R$-charges up to mixing the $R$-symmetry with
$U(1)_A$ (which acts trivially on 4d fields).  As summarized in \autoref{tab:symm},
symmetries are identical to the previous model, except for the extra symmetry $U(1)_A$.
To eliminate $U(1)_A$ we turn on a mass parameter~$\mu$ for it, namely add $\pm\mu$ to the
masses of chiral multiplets as indicated by the markings $+\mu$ and $-\mu$
in~\eqref{wall-ZZ}.  We then take $\mu\to+\infty$.  We will find
that the $S^3_b$ partition function reduces to contributions from
the neighborhood of a point $\tau_\mu=(\mu,\dots,\mu,-\mu,\dots,-\mu)$ of the Coulomb
branch where only half of the chiral multiplets acquire a large mass $2\mu$.  The gauge
group is reduced to $U(N-1)$, and every fundamental chiral multiplet of $\USp(2N-2)$
splits into a fundamental and an antifundamental one, one of which acquires a mass.  The
$\mu\to\pm\infty$ limit is then the $U(N-1)$ 3d theory in~\eqref{wall-ZZ}.  A
separate concern caused by $U(1)_A$ is that the $R$-symmetry of 3d $\Nsusy=2$ can mix
with $U(1)_A$ along the RG flow, and one should perform
$F$-extremization~\cite{Jafferis:2010un} for each value of~$\mu$ to determine the IR
$R$-charges.  Very limited numerical tests suggests that $R$-charges remain bounded, so
that their $\mu$~dependence does not affect the limit.

\subsection{Partition function}
\label{ssec:wall-Z}

\subsubsection{$U(N-1)$ description.}

The coefficients in~\eqref{AGT-ZS3-CBZZ} are products of
$1/\Upsilon_b(x)=\Gamma_b(x)\Gamma_b(b+b^{-1}-x)$ which combine with prefactors of the
braiding kernel~\eqref{kernel-expression}.  Toda CFT momenta are converted to gauge theory
parameters $a$ and~$a'$ and hypermultiplet masses~$m_f$ ($f=1,\ldots,2N$) using the
dictionary~\eqref{AGT-dict}, $\alpha_{12} = Q + \sum_{j=1}^{N} \I a_j h_j$ and
$\alpha_{32} = Q - \sum_{j=1}^{N} \I a'_j h_j$.  Denote $m = \sum_{f=1}^{2N} m_f / (2N)$,
write $m_f=m+\hat{m}_f$, shift all $\tau_j$ and use $1/S_b(x)=S_b(b+b^{-1}-x)$ to find
what partition function to expect given our Toda CFT results:
\begin{equation}\label{wall-ZZZ}
  \begin{aligned}
    & \vev*{\text{S-duality wall on $S^3_b\subset S^4_b$}} \\
    & \, = Z_{\text{1-loop}}^{\text{half-ellipsoid}}(m,a) Z_{\text{1-loop}}^{\text{half-ellipsoid}}(m,a') Z_{S^3_b}(m,a,a') \\
    & \, =
    \frac{\prod_{f=1}^{2N} \prod_{t=1}^{N} \Gamma_b(\frac{b+b^{-1}}{2}+\I\hat{m}_f+\I m-\I a_t)}
    {\prod_{s\neq t}^{N} \Gamma_b(\I a_t-\I a_s)}
    \frac{\prod_{f=1}^{2N} \prod_{t=1}^{N} \Gamma_b(\frac{b+b^{-1}}{2}+\I\hat{m}_f-\I m-\I a'_t)}
    {\prod_{s\neq t}^{N} \Gamma_b(\I a'_t-\I a'_s)}
    \\
    & \quad \times
    \int \frac{\dd[^{N-1}]{\tau}}{\prod_{i\neq j}^{N-1} S_b(\I\tau_i-\I\tau_j)}
    \prod_{j=1}^{N-1} \biggl[
    \prod_{f=1}^{2N} S_b\biggl(\frac{b+b^{-1}}{4}-\I\hat{m}_f+\I\tau_j\biggr)
    \\
    & \qquad\qquad\quad \times
    \prod_{s=1}^{N} S_b\biggl(\frac{b+b^{-1}}{4}-\I m+\I a_s-\I\tau_j\biggr)
    \prod_{s=1}^{N} S_b\biggl(\frac{b+b^{-1}}{4}+\I m+\I a'_s-\I\tau_j\biggr)
    \biggr]
    \,.
  \end{aligned}
\end{equation}
Reassuringly, despite the asymmetry between masses $m_1,\ldots,m_N$ and
$m_{N+1},\ldots,m_{2N}$ in Toda CFT expressions, the final expression is invariant under
permutations of these masses.

The first line in~\eqref{wall-ZZZ} gives perfect candidates for the one-loop determinant
of a hypermultiplet on a half-ellipsoid and that of a vector muliplet,
\begin{equation}
  Z_{\text{1-loop, hyper}}^{\text{half-ellipsoid}}(m) = \Gamma_b\Bigl(\frac{b+b^{-1}}{2}+\I m\Bigr)
  \qquad
  Z_{\text{1-loop, vector}}^{\text{half-ellipsoid}}(a) = \prod_{e\in\{\text{roots}\}} \frac{1}{\Gamma_b\bigl(\I\vev{e|a}\bigr)}
\end{equation}
where $\vev{e|a}$ is the usual scalar product of roots with elements of the Cartan
algebra.  These candidates appear to be consistent with results on the full ellipsoid:
indeed,
\begin{align}
  Z_{\text{1-loop, hyper}}^{S^4_b}(m)
  & =
  Z_{\text{1-loop, hyper}}^{\text{half-}S^4_b}(m)
  Z_{\text{1-loop, hyper}}^{\text{half-}S^4_b}(-m)
  \\
  Z_{\text{1-loop, vector}}^{S^4_b}(a)
  & = Z_{\text{1-loop, vector}}^{\text{half-}S^4_b}(a) Z_{\text{1-loop, vector}}^{\text{half-}S^4_b}(-a)\Bigm/Z_{\text{1-loop, vector}}^{S^3_b}(a) \,.
\end{align}
More precisely, the vector multiplet one-loop determinant is given here for the case of
Neumann boundary conditions (namely gauge transformations are not frozen at the boundary).
The need to divide by the one-loop determinant of a vector multiplet on~$S^3_b$ is not
surprising since we would otherwise be overcounting degrees of freedom.

We are left with the task of finding a 3d $\Nsusy=2$ gauge theory whose $S^3_b$~partition
function is the integral in~\eqref{wall-ZZZ}.  Such partition functions are known through
supersymmetric localization~\cite{Jafferis:2010un,Hama:2010av,Hama:2011ea}: the path
integral is localized to field configurations where a real vector multiplet scalar takes
an arbitrary constant value, which can be reduced to the Cartan algebra of the gauge
group~$G$ by a gauge transformation.  The partition function takes the form
\begin{equation}\label{wall-ZS3b-loc}
  Z_{S^3_b}(\underline{f})
  = \int_{-\infty}^{\infty} \prod_{j=1}^{\rank G} \Bigl[ \dd{\tau_j} e^{\pi\I k \tau_j^2} e^{-2\pi\lambda \tau_j} \Bigr]
  \frac{\prod_{I} \prod_{w_I\in\repr_I} S_b(\frac{1}{2}(b+b^{-1})r_I+\I\vev{w_I|\tau}+\I m_I)}
  {\prod_{e\in\{\text{roots}\}} S_b\bigl(\I\vev{e|\tau}\bigr)}
  \,.
\end{equation}
where the exponentials are classical values of the action, with $k$~the Chern--Simons
level (one per simple factor of~$G$) and $\lambda$~the Fayet--Iliopoulos parameter (one
per abelian factor of~$G$), the product over roots~$e$ of~$G$ is the one-loop contribution
of the vector multiplet, and finally each chiral multiplet transforming in a
representation~$\repr_I$ of~$G$ contributes a product over weights~$w_I$ including
multiplicity (we write $w_I\in\repr_I$ for lack of a better notation), which involves the
$R$-charge $r_I$ and mass~$m_I$ of the chiral.  The integration contour of each $\tau_j$
agrees with $\mathbb{R}$ away from a compact set and is chosen so that for each~$S_b$ all
the poles are above the contour or all below.  Note that the vector multiplet contribution
has no pole since
$\bigl(S_b(y) S_b(- y)\bigr)^{-1} = - 4 \sin(\pi b y) \sin(\pi b^{-1} y)$.

We now find what 3d $\Nsusy=2$ theory reproduces~\eqref{wall-ZZZ} as follows.  From the
product of $S_b(\I\tau_i-\I\tau_j)$ which does not depend on masses we deduce that the
gauge group is $U(N-1)$.  All remaining $S_b$~functions are one-loop determinants of
chiral multiplets with canonical $R$-charge~$\frac{1}{2}$ (since the arguments take the
form $\frac{1}{4}(b+b^{-1})+\cdots$).  The multiplets are $2N$ fundamentals $q_{f}$ of
$U(N-1)$ with masses $-\hat{m}_f$ for $f=1,\ldots,2N$, and $2N$~antifundamentals
$\anti{q}_s$ and $\anti{q}'_{s}$ with masses $a_s-m$ and $a'_s+m$ for $s=1,\ldots,N$.
These masses and $R$-charges are consistent with coupling the 3d and 4d matter multiplets
along the defect using the cubic superpotential
\begin{equation}\label{wall-supo-no-monopole}
  W_{\text{cubic}} = \sum_{f=1}^{2N} \sum_{s=1}^{N} \Bigl( \Phi_{fs}\bigr|_{\text{3d}} \anti{q}_{s} q_{f} + \Phi'_{fs}\bigr|_{\text{3d}} \anti{q}'_{s} q_{f} \Bigr) \,.
\end{equation}
Here $\Phi|_{\text{3d}}$ and~$\Phi'|_{\text{3d}}$ are restrictions of 4d hypermultiplets
at the interface; more precisely they are the 3d $\Nsusy=2$ chiral multiplet whose bottom
component is the 3d restriction of one complex scalar in the hypermultiplets.  The
$R$-charge of these fields originating from 4d must be an integer since the 3d $U(1)_R$
symmetry is embedded in the non-Abelian $SU(2)_R$.  It must be precisely~$1$ because in
flat space the scalars in 4d hypermultiplets transform in a doublet of $SU(2)_R$.
Therefore every term in the superpotential has $R$-charge~$2$.  It is immediate to check
that other charges sum to zero for each term: $(\hat{m}_f+m-a_s)+(a_s-m)+(-\hat{m}_f)=0$
and $(\hat{m}_f-m-a'_s)+(a'_s+m)+(-\hat{m}_f)=0$.

As discussed in \autoref{ssec:wall-theory} the superpotential identifies $SU(2N)$
symmetries of the 3d and 4d theories, and 3d $SU(N)$ symmetries to 4d gauge symmetries.
It also breaks some $U(1)$ symmetries, but leaves four:
\begin{itemize}
\item $U(1)_R$, whose effect on the arguments of $S_b$ functions is to include a term
  $\frac{1}{4}(b+b^{-1})$,
\item $U(1)_1$, whose mass parameter~$m$ we see in masses $a_s-m$ and $a'_s+m$ of
  antifundamental chiral multiplets,
\item $U(1)_B$, whose mass parameter would appear with the same sign in all $S_b$
  functions,
\item $U(1)_T$, whose mass parameter (the FI parameter) would appear as an exponential
  contribution $e^{-2\pi\lambda\sum_j\tau_j}$.
\end{itemize}
We thus need to add to the superpotential some terms that break $U(1)_B$ and $U(1)_T$.  It
turns out that the monopole operators $V_{\pm}$ do the trick, giving the
superpotential~\eqref{intro-supo}.

As explained at the end of \autoref{ssec:AGT-Toda}, if instanton partition functions
(conformal blocks) are normalized to have a leading term $x^{\cdots}$ rather than
$(-x)^{\cdots}$, then the S-duality kernel is changed by phases~\eqref{AGT-phase}
$\exp\bigl(\epsilon\I\pi\bigl[\frac{1}{2}\sum_j a_j^2+\frac{1}{2}\sum_j
a_j'^2\bigr]\bigr)$ depending on the half-plane ($\epsilon=\pm 1$ is the sign of $\Im x$).
These phases are reproduced by a Chern--Simons term of level~$\frac{1}{2}$ for 3d
restrictions of the 4d gauge $SU(N)$ fields.

\subsubsection{$\USp(2N-2)$ description.}

We now consider the $\USp(2N-2)$ theory in~\eqref{wall-ZZ} and explain its partition
function has the $U(N-1)$ one as a limit, up to a divergent factor
$e^{N(N-1)\pi(b+b^{-1})\mu}$ omitted here.  We wish to take $\mu\to\infty$ in the
$S^3_b$~partition function
\begin{equation}\label{wall-uspZ}
  \begin{aligned}
    Z_{S^3_b}
    & =
    \int \dd[^{N-1}]{\tau}
    \prod_{\pm} \Biggl\{
    \frac{\prod_{j=1}^{N-1} \prod_{f=1}^{2N} S_b(\frac{b+b^{-1}}{4}-\I\hat{m}_f-\I\mu\pm\I\tau_j)}
    { \prod_{i\leq j}^{N-1} S_b\bigl(\pm(\I\tau_i+\I\tau_j)\bigr) \prod_{i<j}^{N-1} S_b\bigl(\pm(\I\tau_i-\I\tau_j)\bigr) }
    \\
    & \quad \times
    \prod_{j=1}^{N-1} \prod_{s=1}^{N}
    \biggl[ S_b\Bigl(\frac{b+b^{-1}}{4}-\I m+\I a_s+\I\mu\pm\I\tau_j\Bigr)
    S_b\Bigl(\frac{b+b^{-1}}{4}+\I m+\I a'_s+\I\mu\pm\I\tau_j\Bigr) \biggr]
    \Biggr\}
    \,.
  \end{aligned}
\end{equation}
From the asymptotics of~$\Gamma_b$ \cite[Proposition 8.11]{Spreafico:2009ZZ} we work out
that for $\abs{\chi}\to\pm\infty$ away from the imaginary axis,
\begin{equation}\label{wall-usp-Sblim}
  S_b(A+\I\chi) S_b(B-\I\chi)
  \sim e^{- \pi (b+b^{-1}-A-B) \chi\sign(\Re\chi)} e^{\sign(\Re\chi) \frac{\I\pi}{2} \left( (\frac{1}{2}(b+b^{-1})-B)^2 - (\frac{1}{2}(b+b^{-1})-A)^2\right)} \,.
\end{equation}
We will always take $\chi$~real so $\chi\sign(\Re\chi)=\abs{\chi}$.

Let us apply~\eqref{wall-usp-Sblim} to pairs of $S_b$~functions in~\eqref{wall-uspZ} which
have opposite dependence on $\mu$ and~$\tau_j$, taking $A$ and $B$ to be everything apart
from $\mu$ and~$\tau_j$.  We ignore for now factors that are uniformly bounded functions
of $\mu$ and~$\tau_j$ (and have uniformly bounded inverse), and will denote them by
$(\text{finite})$.  This allows us to keep only the first exponential
in~\eqref{wall-usp-Sblim}.  In fact,
$S_b(A+\I\chi) S_b(B-\I\chi) = e^{- \pi (b+b^{-1}-A-B) \abs{\chi}} (\text{finite})$.  The
integrand becomes
\begin{equation}
  (\text{finite})
  \exp\Biggl( \pi(b+b^{-1}) \biggl(
    -N \sum_{j=1}^{N-1} \Bigl( \abs{\mu-\tau_j} + \abs{\mu+\tau_j} \Bigr)
    + \sum_{i\leq j}^{N-1} \abs{\tau_i+\tau_j}
    + \sum_{i<j}^{N-1} \abs{\tau_i-\tau_j}
  \biggr)\Biggr) \,.
\end{equation}
Now $\abs{\mu-\tau_j} + \abs{\mu+\tau_j} = \max(2\abs{\mu},2\abs{\tau_j})$ and
$\abs{\tau_i-\tau_j} + \abs{\tau_i+\tau_j} = \max(2\abs{\tau_i},2\abs{\tau_j})$.  Sorting
the parameters as
$\abs{\tau_1}<\ldots<\abs{\tau_I}<\abs{\mu}<\abs{\tau_{I+1}}<\cdots<\abs{\tau_{N-1}}$, the
exponential is
\begin{equation}
  \exp\Biggl(\pi(b+b^{-1}) \biggl(
  -N(N-1)\mu
  -2 \sum_{j=1}^{I} j \bigl( \abs{\mu} - \abs{\tau_j} \bigr)
  -2 \sum_{j=I+1}^{N-1} (N-j) \bigl( \abs{\tau_j} - \abs{\mu} \bigr)
  \biggr)\Biggr) \,.
\end{equation}
The second and third terms are negative.  Hence the dominant contribution to the
integral~\eqref{wall-uspZ} as $\mu\to\infty$ is when $\abs{\tau_j}-\abs{\mu}$ are finite,
and away from these regions the integrand decays exponentially.

We can now go back and keep finite factors when expanding the integrand in the region
$\abs{\tau_j}\sim\abs{\mu}$.  Given symmetries under $\tau_j\to-\tau_j$, we focus on the
case $\tau_j=\mu+\hat{\tau_j}$ with $\hat{\tau_j}$ finite.  Half of the $S_b$ factors
remain finite and form the partition function of the $U(N-1)$ theory, while the other half
can be paired and turned into exponentials through~\eqref{wall-usp-Sblim}.  One gets
\begin{equation}\label{wall-usp-to-u}
  \begin{aligned}
    & \lim_{\mu\to\pm\infty} \Bigl[ E(\mu,b,m,\hat{m},a,a') Z_{S^3_b} \bigl(\text{$\USp(2N-2)$ theory}\bigr) \Bigr]
    \\
    & =
    \int \dd[^{N-1}]{\hat{\tau}}
    \Biggl\{
    \frac{\prod_{j=1}^{N-1} \prod_{f=1}^{2N} S_b(\frac{b+b^{-1}}{4}-\I\hat{m}_f+\I\hat{\tau}_j)}
    { \prod_{\pm} \prod_{i<j}^{N-1} S_b\bigl(\pm(\I\hat{\tau}_i-\I\hat{\tau}_j)\bigr) }
    \\
    & \qquad \times
    \prod_{j=1}^{N-1} \prod_{s=1}^{N}
    \biggl[ S_b\Bigl(\frac{b+b^{-1}}{4}-\I m+\I a_s-\I\hat{\tau}_j\Bigr)
    S_b\Bigl(\frac{b+b^{-1}}{4}+\I m+\I a'_s-\I\hat{\tau}_j\Bigr) \biggr]
    \Biggr\}
    \\
    & = Z_{S^3_b} \bigl(\text{$U(N-1)$ theory}\bigr)
    \,.
  \end{aligned}
\end{equation}
where
$E = \Bigl[ \frac{1}{2} e^{N\pi(b+b^{-1})\abs{\mu}}
e^{\sign(\mu)\frac{\I\pi}{2}\bigl(\sum_{f=1}^{2N} \hat{m}_f^2 - \sum_{s=1}^{N} a_s^2 -
  \sum_{s=1}^{N} a_s'^2 - 2N m^2\bigr)} \Bigr]^{N-1}$.  As announced, the partition
function of the $U(N-1)$ theory is a limit of the partition function of the $\USp(2N-2)$
theory.  See~\cite{Benini:2017dud} for the analogous statement for the theories
themselves, although one may want to keep track of the factor~$E$, which consists of
Chern--Simons terms for all flavour symmetry groups of the 3d theories.

\subsection{Discrete symmetries and dualities}
\label{ssec:wall-duality}

We now have all the necessary tools to study symmetries, from the point of view of the
duality wall, of its 3d gauge theory description, and of the Toda CFT\@.  Symmetries of Toda
CFT correlators (hence of their braiding kernel) may be the least familiar to the reader, so
we will describe them and give their gauge theory interpretation.

Any vertex operator~$\widehat{V}_{\alpha}$ is invariant under Weyl transformations, namely
permutations of the components $\vev{\alpha-Q,h_s}$.  Applying this symmetry
to~$\widehat{V}_{\alpha_1}$ permutes the first~$N$ masses, and applying it
to~$\widehat{V}_{\alpha_3}$ permutes the last~$N$.  These are manifest symmetries of the
S-duality wall and of its 3d gauge theory description.  In fact, permuting all of the
$2N$~masses is also a symmetry, but the Toda CFT description does not make all
permutations manifest.

Conformal blocks are invariant under conjugating all momenta, which maps (up to a Weyl
transformation) $\alpha\to 2Q-\alpha$ and $\kappa h_1\to (N(b+b^{-1})-\kappa)h_1$ hence
flips the sign of all
$m_f$, $a_s$ and~$a'_s$.  This is the effect of charge conjugation in the 4d theories on
both sides of the wall, which is an expected symmetry of the S-duality wall.  However, the
effect is harder to describe in the 3d gauge theory description, because this theory is
chiral, hence not invariant a priori under charge conjugation.  Correspondingly, the
explicit form of the braiding kernel does not appear invariant under charge conjugation.
Most of this section will be spent proving the invariance, which will turn out to be a
limit of $\USp$-type Seiberg duality of 4d $\Nsusy=1$ indices.  Before proceeding let us
describe two more symmetries.

Conformal blocks are also invariant under some permutations of the operators.  In
particular, the braiding kernel is thus invariant under a permutation obtained by rotating
the diagrams by $180^{\circ}$ in the plane:
\begin{equation}
  \Braiding{}{\alpha_{12},\alpha_{32}}{\kappa_4 h_1&\kappa_2 h_1\\\alpha_3&\alpha_1}
  = \Braiding{}{2Q-\alpha_{12},2Q-\alpha_{32}}{\kappa_2 h_1&\kappa_4 h_1\\\alpha_1&\alpha_3}
  \text{ due to}
  \smash{
  \mathtikz[x=1em,y=1ex,thick]{
    \draw[->-=.55] (0,0) -- (2,0);
    \draw[->-=.55] (6,0) -- (4,0);
    \draw[->-=.55] (4,0) -- (2,0);
    \draw[->-=.55] (2,-4) -- (2,0);
    \draw[->-=.55] (4,4) -- (4,0);
    \node at (0.3,1.5) {$\alpha_3$};
    \node at (5.7,-1.5) {$\alpha_1$};
    \node at (3,1.5) {$\alpha_{12}$};
    \node at (5.3,3) {$\kappa_2 h_1$};
    \node at (0.7,-3) {$\kappa_4 h_1$};
    \draw[->] (-1,-1) to [bend left=90] (7,1);
    \path (7,1) to [bend left=90] (-1,-1);
  }
  }
\end{equation}
The map $\alpha_{12}\to 2Q-\alpha_{12}$ is due to the arrow being reversed by the
rotation.  Composing with conjugation of all momenta yields the transformation
$\alpha_1\leftrightarrow 2Q-\alpha_3$ and $\kappa_2\leftrightarrow N(b+b^{-1})-\kappa_4$, which
given the dictionary~\eqref{AGT-dict} simply exchanges the first~$N$ and the last~$N$
masses.

Lastly, the S-duality wall and its 3d gauge theory description are invariant under
exchanges of the two hemispheres.  This symmetry is manifest in explicit expressions, but
on the Toda side it is not immediately clear why it should hold.  The symmetry maps
$a_s\leftrightarrow a'_s$ and $m\to-m$.  Momenta are mapped as
$\kappa_2\leftrightarrow\kappa_4$ and $\alpha_{12}\leftrightarrow\alpha_{32}$, and the
braiding kernel becomes the kernel for the inverse braiding, which expresses u-channel
blocks in terms of s-channel ones.  In addition to being inverses of each other, these
braiding kernels are actually also equal up to some structure constants.  To prove this,
write the s-channel and u-channel decompositions of the four-point function, then use
braiding kernels to get an expression with a holomorphic s-channel block and an
antiholomorphic u-channel block, and match coefficients of
$\Fblock{\sch}{\alpha_{12}}{}{x} \Fblock{\uch}{\alpha_{32}}{}{\bar{x}}$:
\begin{align}
  \vev{\widehat{V}\widehat{V}\widehat{V}\widehat{V}}
  & = \!\int\!\dd{\alpha_{12}} C^{\sch}\! \Fblock{\sch}{\alpha_{12}}{}{x} \Fblock{\sch}{\alpha_{12}}{}{\bar{x}}
  = \!\int\!\dd{\alpha_{12}} \dd{\alpha_{32}}
    C^{\sch}\! \Fblock{\sch}{\alpha_{12}}{}{x} \Braiding{}{\alpha_{12}\alpha_{32}}{} \Fblock{\uch}{\alpha_{32}}{}{\bar{x}}
  \\
  \vev{\widehat{V}\widehat{V}\widehat{V}\widehat{V}}
  & = \!\int\!\dd{\alpha_{32}} C^{\uch}\! \Fblock{\uch}{\alpha_{32}}{}{x} \Fblock{\uch}{\alpha_{32}}{}{\bar{x}}
  = \!\int\!\dd{\alpha_{12}} \dd{\alpha_{32}}
    C^{\uch} \Braiding{\kappa_2\leftrightarrow\kappa_4}{\alpha_{32}\alpha_{12}}{}
    \Fblock{\sch}{\alpha_{12}}{}{x} \Fblock{\uch}{\alpha_{32}}{}{\bar{x}}
  .
\end{align}
These two decompositions into the basis of functions
$\Fblock{\sch}{\alpha_{12}}{}{x} \Fblock{\uch}{\alpha_{32}}{}{\bar{x}}$ must coincide hence
$C^{\sch} \Braiding{}{\alpha_{12}\alpha_{32}}{} = C^{\uch} \Braiding{\kappa_2\leftrightarrow\kappa_4}{\alpha_{32}\alpha_{12}}{}$.
The structure constants $C^{\sch}$ and~$C^{\uch}$ turn out to cancel with the prefactors relating the braiding
kernel to the S-duality wall partition function so that we get an equality of wall partition functions.

Now that we have described all manifest symmetries of the explicit braiding
kernel~\eqref{kernel-expression}, we must tackle invariance under charge conjugation.

We will use identities of hyperbolic hypergeometric integrals~\cite{Rains:2006ZZ}.  The
hyperbolic Gamma function~$\Gamma^{(2)}_{h}$ of that paper reduces to our~$S_b$ upon
taking $\omega_1/\omega_2=b^2$.  For definiteness, we take $\omega_1=\I b$ and
$\omega_2=\I/b$ and note $S_b(x)=\Gamma^{(2)}_{h}(\I x;\I b,\I/b)$.  The $BC_n$ hyperbolic
hypergeometric integral is defined by (to avoid factors of~$\I$ we let $\mu_r=\I\nu_r$)
\begin{equation}\label{wall-dual-IBC}
  I_{BC_n;h}^{(m)}(\nu_0,\ldots,\nu_{2m+2n+3})
  = \frac{1}{2^n n!} \int_{\mathbb{R}^n} \dd[^n]{x} \prod_{\pm}
  \frac{\prod_{i=1}^n \prod_{r=0}^{2m+2n+3} S_b(\nu_r\pm\I x_i)}
  {\prod_{i\leq j}^n S_b\bigl(\pm(\I x_i+\I x_j)\bigr) \prod_{i<j}^n S_b\bigl(\pm(\I x_i-\I x_j)\bigr)}
\end{equation}
for $\sum_{r=0}^{2m+2n+3} \nu_r=(m+1)(b+b^{-1})$.  Corollary 4.2
of~\cite{Rains:2006ZZ} states the invariance under $m\leftrightarrow n$ and
$\nu\to \frac{1}{2}(b+b^{-1})-\nu$:
\begin{equation}\label{wall-dual-Rains}
  I_{BC_n;h}^{(m)}\bigl(\nu_0,\ldots,\nu_{2m+2n+3}\bigr)
  = \biggl[\prod_{r<s}^{2m+2n+3} \!\!S_b(\nu_r+\nu_s)\biggr]
  I_{BC_m;h}^{(n)}\bigl(\tfrac{b+b^{-1}}{2}-\nu_0,\tfrac{b+b^{-1}}{2}-\nu_1,\ldots\bigr) \,.
\end{equation}
The $S^3_b$~partition function~\eqref{wall-uspZ} of the $\USp(2N-2)$ SQCD with $4N$ chiral
multiplets which we studied earlier is such a hyperbolic hypergeometric integral, with
$m=n=N-1$ and $4N$~parameters $\nu_r$ summing to $N(b+b^{-1})$:
\begin{equation}
  Z_{S^3_b} = I_{BC_{N-1};h}^{(N-1)}
  \begin{aligned}[t]
    & \Bigl(
    \tfrac{b+b^{-1}}{4}-\I\hat{m}_f-\I\mu \ (f=1,\ldots,2N),\\
    & \quad \tfrac{b+b^{-1}}{4}-\I m+\I a_s+\I\mu,
    \tfrac{b+b^{-1}}{4}+\I m+\I a'_s+\I\mu
    \ (s=1,\ldots,N)
    \Bigr)
    \,.
  \end{aligned}
\end{equation}
The identity~\eqref{wall-dual-Rains} states that $Z_{S^3_b}$ is invariant up to a factor
$\prod_{r<s} S_b(\nu_r+\nu_s)$ under changing the signs of all $\mu$, $\hat{m}_f$, $m$,
$a_s$, $a'_s$.

We now take the limit $\mu\to\pm\infty$ in~\eqref{wall-dual-Rains}, after multiplying by
$E(\mu)$ and $E(-\mu)$ respectively as explained in~\eqref{wall-usp-to-u}.  On each side,
we obtain partition functions of $U(N-1)$ theories, and the product of $S_b(\nu_r+\nu_s)$
reads
\begin{equation}\label{wall-dual-prefactor}
  \begin{aligned}
    \prod_{r<s}^{4N-1} S_b(\nu_r+\nu_s)
    & =
    \prod_{f=1}^{2N} \prod_{s=1}^{N} \biggl[
    S_b\biggl(\frac{b+b^{-1}}{2}-\I\hat{m}_f-\I m+\I a_s\biggr) S_b\biggl(\frac{b+b^{-1}}{2}-\I\hat{m}_f+\I m+\I a'_s\biggr) \biggr]
    \\
    & \qquad\times
    e^{-\I\pi\bigl(
        \sum_{f=1}^{2N}\hat{m}_f^2
        -\sum_{s=1}^{N} a_s^2
        -\sum_{s=1}^{N} a_s'^2
        -2Nm^2
      \bigr)(N-1)\sign\mu} \,.
  \end{aligned}
\end{equation}
The phases are exactly $E(-\mu)/E(\mu)$ with $E$ defined below~\eqref{wall-usp-to-u}.  The
remaining $S_b$~functions are one-loop
determinants of mesons formed as the product of a fundamental and an antifundamental
chiral multiplet under $U(N-1)$.  The full $S^4_b$~partition function~\eqref{wall-ZZZ} of
the 4d/3d coupled system also includes one-loop determinants of hypermultiplets on
half-ellipsoids, which are $\Gamma_b$ functions with the same arguments as the $S_b$
functions in~\eqref{wall-dual-prefactor} up to signs.  Since
\begin{equation}
  S_b\biggl(\frac{b+b^{-1}}{2}+x\biggr)=
  \frac{\Gamma_b\bigl(\frac{b+b^{-1}}{2}+x\bigr)}{\Gamma_b\bigl(\frac{b+b^{-1}}{2}-x\bigr)},
\end{equation}
the $S_b$~functions coming from mesons in the 3d duality convert between
$\Gamma_b\bigl(\frac{b+b^{-1}}{2}\pm x\bigr)$.  This is fully consistent with 4d charge conjugation.
Identical calculations show that the braiding kernel is invariant under Toda CFT charge
conjugation.

The identity~\eqref{wall-dual-Rains} was proven in~\cite{Rains:2006ZZ} as a hyperbolic
limit of an elliptic hypergeometric integral identity.  In physics terms, the elliptic
identity states that two Seiberg-dual 4d $\Nsusy=1$ theories with $\USp(2N-2)$ gauge group
and $4N$~fundamental chiral multiplets have the same $S^3_b\times S^1$ partition function
(supersymmetric index).  Note that while the dual of 4d $\Nsusy=1$ $\USp(2N_c)$ SQCD with
$2N_f$~fundamental chiral multiplets has $\widetilde{N}_c = N_f-N_c-2$ colors, the
analoguous 3d $\Nsusy=2$ Aharony duality has $\widetilde{N}_c = N_f-N_c-1$ colors
instead~\cite{Aharony:1997gp}.  For our case $N_c=N-1$ and $N_f=2N$ the 4d theory is
self-dual but the 3d theory is not.  On the other hand, Aharony duality can be retrieved
as a limit of~\eqref{wall-dual-Rains} when $\nu_{2m+2n+2},-\nu_{2m+2n+3}\to\I\infty$.
This is physically interpreted as dimensional reduction of 4d $\USp$ dualities down to 3d
following~\cite{Aharony:2013dha,Aharony:2013kma}.

Let us finish this section with a mention of $N=2$.  Besides the $\USp$-type Seiberg
duality there is also an $SU$-type Seiberg duality in 4d in that case, thanks to the
accidental
isomorphism $\USp(2)=SU(2)$.  The map of parameters for the $SU(2)$ Seiberg duality depends
on a split of the $8$~chiral multiplets into $4$ ``fundamental'' and $4$
``antifundamental'' ones.  After descending to 3d $\Nsusy=2$ partition functions,
\begin{align}\label{wall-dual-SU2}
  & I_{BC_1;h}^{(1)}\bigl(\nu_0,\ldots,\nu_7\bigr)
  \\\nonumber
  & = \prod_{r=0}^{3}\prod_{s=4}^{7}\Bigl[ S_b(\nu_r+\nu_s)\Bigr]
    I_{BC_1;h}^{(1)}
    \Bigl(\tfrac{\nu_0+\nu_1}{2}\pm\tfrac{\nu_2-\nu_3}{2},
    \tfrac{\nu_2+\nu_3}{2}\pm\tfrac{\nu_0-\nu_1}{2},
    \tfrac{\nu_4+\nu_5}{2}\pm\tfrac{\nu_6-\nu_7}{2},
    \tfrac{\nu_6+\nu_7}{2}\pm\tfrac{\nu_4-\nu_5}{2}\Bigr) \,.
\end{align}
As for general~$N$, we are interested in a limit where the~$\nu$ go to $\pm\I\infty$ (half
with each sign).  We can in particular choose
$+\nu_0, -\nu_1, -\nu_2, -\nu_3, +\nu_4, +\nu_5, +\nu_6, -\nu_7\sim+\I\mu$.  Then most
parameters in the dual~\eqref{wall-dual-SU2} remain finite: only the first and the last
depend on~$\mu$.  The limit $\mu\to\infty$ then takes the $S^3_b$~partition function of
$SU(2)$ SQCD with $8$~doublets to that of $SU(2)$ SQCD with $6$~doublets.  This reproduces
the 3d $\Nsusy=2$ description found in~\cite{Teschner:2012em} for the S-duality domain
wall of 4d $\Nsusy=2$ $SU(2)$ SQCD with $4$~flavours.

For $N=2$, our description of the domain wall as a $U(N-1)=U(1)$ theory, namely 3d
$\Nsusy=2$ SQED with $N_f=4$ flavours and a monopole superpotential was also found to be
dual to $SU(2)$ with $6$ doublets in~\cite{Dimofte:2012pd} by realizing both from
dimensional reduction of a 4d theory with $E_7$ symmetry.  It would be interesting to see
if a similar construction could be performed for $N>2$.

\section{Conclusions}
\label{sec:conclusion}

In this work we have determined the integral kernel~\eqref{kernel-expression} which braids
two semi-degenerate vertex operators of the Toda CFT\@.  Through the AGT relation, we have
deduced the ellipsoid expectation value of an S-duality domain wall in 4d $\Nsusy=2$
$SU(N)$ SQCD with $2N$~flavours.  We have then described the wall by coupling 3d
$\Nsusy=2$ $U(N-1)$ SQCD with $2N+2N$ chiral multiplets and a monopole superpotential on
the wall to the 4d theories on both sides of the wall.

The shift relations found in \autoref{ssec:kernel-shift} do not appear sufficient to prove
that the braiding kernel is correct.  An obvious question would be to fill in this gap by
checking additional Moore--Seiberg relations.  It should be possible to extract
Racah--Wigner coefficients for the modular double of $U_q(sl_N)$, as was done for $N=2$
in \cite{Ponsot:1999uf,Ponsot:2000mt},
and to get 6j symbols of $U_q(sl_N)$ by taking a limit of
degenerate momenta, thus generalizing~\cite{Nawata:2013ooa} from four
symmetric to two symmetric representations of $U_q(sl_N)$.

It would be valuable to evaluate one-loop determinants of hypermultiplets and vector
multiplets on the half-ellipsoid with appropriate boundary conditions, and clarify whether
the result indeed consists of half of the~$\Gamma_b$ factors in the full-ellipsoid
results.

S-duality is expected to map Wilson loops to 't~Hooft loops.  Expectation values of Wilson
loop and 't~Hooft loop observables on the ellipsoid are known
exactly~\cite{Pestun:2007rz,Gomis:2011pf} and we now know the explicit S-duality kernel.
Conjugating the Wilson loop by this kernel should thus yield the 't~Hooft loop, namely if
one considers the 4d theory on~$S^4_b$ with two S-duality walls near the equator and a
Wilson loop in between, then the collision limit should yield a 't Hooft loop.  In fact, a
preliminary question is to understand how the collision of two domain walls which perform
S-duality and its inverse yields a trivial operator.  This annihilation should involve
appropriate 3d dualities, as happens in one dimension higher in~\cite{Gaiotto:2015una}.

The S-duality wall generalizes straightforwardly to the case where part (or all) of the
$SU(2N)$ flavour
symmetry shared by the two 4d theories is gauged by 4d vector multiplets.  Class~S
theories, constructed by twisted dimensional reduction of the 6d $(2,0)$ $SU(N)$
superconformal theory on a punctured Riemann surface~$\Sigma$, provide interesting
examples, for instance linear quivers of $SU(N)$ gauge groups.  The 3d description of an
S-duality wall in such a quiver is again 3d $\Nsusy=2$ $U(N-1)$ SQCD coupled with 4d
fields through cubic superpotentials:
\begin{equation}
  \quiver{
    \node (LLLL)[flavor-group]                    {$N$};
    \node (LLL) [color-group, right=1.5em of LLLL]{$N$};
    \node (LL)  [color-group, right=1.5em of LLL] {$N$};
    \node (L)   [color-group, right=1.5em of LL]  {$N$};
    \node (1)   [color-group, right=1.5em of L] {$\scriptstyle U(N-1)$};
    \node (R)   [color-group, right=1.5em of 1] {$N$};
    \node (RR)  [flavor-group, right=1.5em of R] {$N$};
    \node (A)   [color-group, above of=1] {$N$};
    \node (B)   [color-group, below of=1] {$N$};
    \draw[->-=.55] (L) -- (1);
    \draw[->-=.55] (R) -- (1);
    \draw[->-=.55] (1) -- (A);
    \draw[->-=.55] (1) -- (B);
    \draw (LLLL) -- (LLL) -- (LL) -- (L) -- (A) -- (R);
    \draw (RR) -- (R) -- (B) -- (L);
  }
\end{equation}
Quiver gauge theories open up the possibility of colliding S-duality domain walls, where
S-duality acts on different gauge groups~$SU(N)$.  When the groups are separated, the 3d
description is obvious (on the Toda side the braiding operations commute):
\begin{equation}
  \quiver{
    \node (0)   [flavor-group] {$N$};
    \node (1)   [color-group, right=1.5em of 0] {$N$};
    \node (d2)  [color-group, right=1.5em of 1] {$\scriptstyle U(N-1)$};
    \node (2)   [color-group, above of=d2] {$N$};
    \node (2')  [color-group, below of=d2] {$N$};
    \node (3)   [color-group, right=1.5em of d2] {$N$};
    \node (d4)  [color-group, right=1.5em of 3] {$\scriptstyle U(N-1)$};
    \node (4)   [color-group, above of=d4] {$N$};
    \node (4')  [color-group, below of=d4] {$N$};
    \node (5)   [color-group, right=1.5em of d4] {$N$};
    \node (6)   [flavor-group, right=1.5em of 5] {$N$};
    \draw (0) -- (1) -- (2) -- (3) -- (4) -- (5) -- (6);
    \draw (1) -- (2') -- (3);
    \draw (3) -- (4') -- (5);
    \draw[->-=.55] (1) -- (d2);
    \draw[->-=.55] (3) -- (d2);
    \draw[->-=.55] (d2) -- (2);
    \draw[->-=.55] (d2) -- (2');
    \draw[->-=.55] (3) -- (d4);
    \draw[->-=.55] (5) -- (d4);
    \draw[->-=.55] (d4) -- (4);
    \draw[->-=.55] (d4) -- (4');
  }
\end{equation}
When the two groups share a bifundamental hypermultiplet, the duality walls do not commute
and we propose the following descriptions for the two possible orderings:
\begin{equation}
  \quiver{
    \node (1)   [flavor-group] {$N$};
    \node (d2)  [color-group, right=1.5em of 1] {$\scriptstyle U(N-1)$};
    \node (2)   [color-group, above of=d2] {$N$};
    \node (2')  [color-group, below of=d2] {$N$};
    \node (d3)  [color-group, right=1.5em of d2] {$\scriptstyle U(N-1)$};
    \node (3)   [color-group, above of=d3] {$N$};
    \node (3')  [color-group, below of=d3] {$N$};
    \node (4)   [flavor-group, right=1.5em of d3] {$N$};
    \draw (1) -- (2) -- (3) -- (4);
    \draw (1) -- (2') -- (3') -- (4);
    \draw[->-=.55] (1) -- (d2);
    \draw[->-=.55] (d2) -- (2);
    \draw[->-=.55] (d2) -- (2');
    \draw[->-=.55] (4) -- (d3);
    \draw[->-=.55] (d3) -- (3);
    \draw[->-=.55] (d3) -- (3');
    \draw[->-=.55] (3) -- (d2);
    \draw[->-=.55] (2') -- (d3);
  }
  \quad\text{vs}\quad
  \quiver{
    \node (1)   [flavor-group] {$N$};
    \node (d2)  [color-group, right=1.5em of 1] {$\scriptstyle U(N-1)$};
    \node (2)   [color-group, above of=d2] {$N$};
    \node (2')  [color-group, below of=d2] {$N$};
    \node (d3)  [color-group, right=1.5em of d2] {$\scriptstyle U(N-1)$};
    \node (3)   [color-group, above of=d3] {$N$};
    \node (3')  [color-group, below of=d3] {$N$};
    \node (4)   [flavor-group, right=1.5em of d3] {$N$};
    \draw (1) -- (2) -- (3) -- (4);
    \draw (1) -- (2') -- (3') -- (4);
    \draw[->-=.55] (1) -- (d2);
    \draw[->-=.55] (d2) -- (2);
    \draw[->-=.55] (d2) -- (2');
    \draw[->-=.55] (4) -- (d3);
    \draw[->-=.55] (d3) -- (3);
    \draw[->-=.55] (d3) -- (3');
    \draw[->-=.55] (3') -- (d2);
    \draw[->-=.55] (2) -- (d3);
  }
\end{equation}
with a cubic superpotential term coupling 3d and 4d fields for each triangle and a quartic
superpotential for 3d chiral multiplets in the central paralellogram, as well as suitable monopole superpotentials.
Such combinations of S-duality walls have been investigated (after version~1 of the present paper) in~\cite{Garozzo:2019xzi}.

From the Toda CFT point of view, each of these products of duality walls corresponds to a
product of two $W_N$ braiding kernels.  It may be interesting to translate Moore--Seiberg
relations of braiding kernels into the gauge theory language and understand how 3d
$\Nsusy=2$ dualities reproduce them.

Note that all of this work focused on gauge theories with a Lagrangian description, or
equivalently Toda CFT correlators with ``enough'' degeneracy.  In particular, we have
avoided the limit $x\to 1$ of 4d $\Nsusy=2$ SQCD, which involves a strongly coupled matter
theory instead of hypermultiplets, coupled to a vector multiplet.  For $N=3$ this theory
includes~$T_3$.  The corresponding crossing symmetry on the Toda CFT side consists of the
fusion of two simple punctures into a less degenerate operator~$\widehat{V}_{\alpha}$.
Conformal blocks in this limit are not uniquely characterized by~$\alpha$ and external
operators, and one needs a label for the continuous multiplicity with
which~$\widehat{V}_{\alpha}$ appears in the fusion of the two full punctures.  These
conformal blocks are eigenfunctions of the square of the braiding kernel, and it is
tempting to try and diagonalize this kernel.  Unfortunately, we only succeeded to tame
multiplicities in the simplest discrete versions of the kernel, and could not generalize.

Another direction worth pursuing is to consider S-duality walls in 4d $\Nsusy=2$ theories
with gauge groups such as $Sp(N)$ (see for instance~\cite{Lee:1997fy}).  Too little is
known at present about braiding kernels of D-type and E-type Toda CFT in order to apply
the techniques used here.  Understanding whether the $U(N-1)$ 3d theory found in this
paper can be derived through brane constructions may help in generalizing to other gauge
groups by orbifolding.

It would also be interesting to check that our gauge theory description of the S-duality
wall gives the correct $S^3_b\times S^1$ index, as was done for $SU(2)$ theories
in~\cite{Gang:2012ff,Koh:2013sra,Gang:2013sqa,Bullimore:2014nla}.

\appendix

\acknowledgments

I thank Francesco Benini, Bertrand Eynard, Yale Fan, Davide Gaiotto,
Igor Klebanov, M\'ark Mezei, Satoshi Nawata, Sara Pasquetti, Silviu
Pufu, Shlomo Razamat, Sylvain Ribault, Nathan Seiberg, Jan Troost, Ran
Yacoby and especially Jaume Gomis for discussions, comments, and
pointers to the literature.

\begingroup\raggedright

\endgroup

\end{document}